\documentclass[prd,aps,superscriptaddress,tightenlines,nofootinbib,floatfix,preprintnumbers,11pt,longbibliography]{revtex4-1}
\usepackage{epsfig}
\usepackage{amsfonts}
\usepackage{amsmath}
\usepackage{amssymb}
\usepackage{dsfont}
\usepackage{hyperref}
\usepackage{xspace}
\usepackage{slashed}
\usepackage{braket}

\hypersetup{
    colorlinks=true,       
    linkcolor=blue,          
    citecolor=blue,        
    filecolor=blue,      
    urlcolor=blue           
}
\usepackage{xcolor}
\usepackage{color}
\usepackage{graphicx}  %
\usepackage{bm}  %
\usepackage{graphics}

\usepackage{multirow}

\usepackage{enumitem}

\newcommand{\dslash}[1]{#1 \llap{/\kern-0.5pt}}
\newcommand{\Dslash}[1]{#1 \llap{/\kern+1.5pt}}
\newcommand{\DDslash}[1]{#1 \llap{/\kern+2.3pt}}
\newcommand{\dslashh}[1]{#1 \llap{/\kern+1pt}}

\newcommand{\bea}{\begin{eqnarray}}
\newcommand{\eea}{\end{eqnarray}}
\newcommand{\be}{\begin{equation}}
\newcommand{\ee}{\end{equation}}
\newcommand{\bma}{\begin{pmatrix}}
\newcommand{\ema}{\end{pmatrix}}

\newcommand{\nuc}[2]{\ensuremath{^{#2}\mathrm{#1}}}

\newcount\Comments  
\newcommand{\kibitz}[2]{\ifnum\Comments=1\textcolor{#1}{#2}\fi}

\begin{document}

\title{
\vspace*{0.5cm}
Neutrinoless Double-Beta Decay: 
\\
A Roadmap for Matching Theory to Experiment
\vspace*{.5cm}
}

\author{Vincenzo Cirigliano}
\affiliation{Institute for Nuclear Theory, University of Washington, Seattle WA 91195-1550, USA}

\author{Zohreh~Davoudi}
\email[Corresponding author: ]{davoudi@umd.edu}
\affiliation{Maryland Center for Fundamental Physics and Department of Physics, University of
Maryland, College Park, MD 20742, USA}

\author{Wouter~Dekens}
\affiliation{Institute for Nuclear Theory, University of Washington, Seattle WA 91195-1550, USA}

\author{Jordy~de~Vries} 
\affiliation{ Institute for Theoretical Physics Amsterdam and Delta Institute for Theoretical Physics, University of Amsterdam, Science Park 904, 1098 XH Amsterdam, The Netherlands}
\affiliation{Nikhef, Theory Group, Science Park 105, 1098 XG, Amsterdam, The Netherlands}

\author{Jonathan~Engel}
\email[Corresponding author: ]{engelj@physics.unc.edu}
\affiliation{Department of Physics and Astronomy, University of North Carolina, Chapel Hill, North Carolina 27516-3255, USA}

\author{Xu~Feng}
\affiliation{School of Physics, Peking University, Beijing 100871, China}
\affiliation{
Collaborative Innovation Center of Quantum Matter, Beijing 100871, China}
\affiliation{
Center for High Energy Physics, Peking University, Beijing 100871, China}

\author{Julia~Gehrlein}
\email[Corresponding author: ]{jgehrlein@bnl.gov}
\affiliation{High Energy Theory Group, Physics Department, Brookhaven National Laboratory, Upton, NY 11973, USA}

\author{Michael~L.~Graesser}
\email[Corresponding author: ]{michaelgraesser@gmail.com}
\affiliation{Theoretical Division, Los Alamos National Laboratory, Los Alamos, 
NM 87545, USA}

\author{Luk\'{a}\v{s}~Gr\'{a}f}
\affiliation{Department of Physics, University of California, Berkeley, CA 94720, USA}
\affiliation{Department of Physics, University of California, San Diego, CA 92093, USA}

\author{Heiko~Hergert}
\email[Corresponding author: ]{hergert@frib.msu.edu}
\affiliation{Facility for Rare Isotope Beams and Department of Physics \& Astronomy, Michigan State University, East Lansing, MI 48824-1321, USA}

\author{Luchang~Jin}
\affiliation{Department of Physics, University of Connecticut, Storrs, CT 06269, USA}
\affiliation{RIKEN-BNL Research Center, Brookhaven National Laboratory, Building 510, Upton, NY 11973, USA}

\author{Emanuele~Mereghetti}
\email[Corresponding author: ]{emereghetti@lanl.gov}
\affiliation{Theoretical Division, Los Alamos National Laboratory, Los Alamos, 
NM 87545, USA}

\author{Amy~Nicholson}
\email[Corresponding author: ]{annichol@email.unc.edu}
\affiliation{Department of Physics and Astronomy, University of North Carolina, Chapel Hill, NC 27516-3255, USA}

\author{Saori~Pastore}
\affiliation{Department of Physics, Washington University in Saint Louis, Saint Louis, MO 63130, USA}
\affiliation{McDonnell Center for the Space Sciences at Washington University in St. Louis, MO 63130, USA}

\author{Michael~J.~Ramsey-Musolf}
\email[Corresponding author: ]{mjrm@sjtu.edu.cn}
\affiliation{Tsung Dao Lee Institute, Shanghai Jiao Tong University, Shanghai 200120 China}
\affiliation{University of Massachusetts, Amherst, MA 01003 USA}

\author{Richard~Ruiz}
\affiliation{Institute of Nuclear Physics -- Polish Academy of Sciences {\rm (IFJ PAN)}, Krak{\'o}w 31-342, Poland}

\author{Martin~Spinrath}
\affiliation{Department of Physics, National Tsing Hua University, Hsinchu, 30013, Taiwan}\affiliation{
Center for Theory and Computation, National Tsing Hua University, Hsinchu, 30013, Taiwan
}

\author{Ubirajara~van~Kolck}
\affiliation{Universit\'e Paris-Saclay, CNRS/IN2P3, IJCLab, 91405 Orsay, France}
\affiliation{Department of Physics, University of Arizona, Tucson, AZ 85721, USA}

\author{Andr\'e~Walker-Loud}
\affiliation{Nuclear Science Division, Lawrence Berkeley National Laboratory, Berkeley, CA 94720, USA}

\newcommand\snowmass{\begin{center}\rule[-0.2in]{\hsize}{0.01in}\\\rule{\hsize}{0.01in}\\
\vskip 0.1in Submitted to the  Proceedings of the U.S. Community Study\\ 
on the Future of Particle Physics (Snowmass 2021)\\ 
\rule{\hsize}{0.01in}\\\rule[+0.2in]{\hsize}{0.01in} \end{center}}

\snowmass

\begin{abstract}
\vspace{1 cm}
\noindent
\textbf{Abstract.} The observation of neutrino oscillations and hence non-zero neutrino masses provided a milestone in the search for physics beyond the Standard Model. But even though we now know that neutrinos are massive, the nature of neutrino masses, i.e., whether they are Dirac or Majorana, remains an open question. A smoking-gun signature of Majorana neutrinos is the observation of neutrinoless double-beta decay, a process that violates the lepton-number conservation of the Standard Model. This white paper focuses on the theoretical aspects of the neutrinoless double-beta decay program and lays out a roadmap for future developments. The roadmap is a multi-scale path starting from high-energy models of neutrinoless double-beta decay all the way to the low-energy nuclear many-body problem that needs to be solved to supplement measurements of the decay rate. The path goes through a systematic effective-field-theory description of the underlying processes at various scales and needs to be supplemented by lattice quantum chromodynamics input. 

\vspace{.1cm}
\noindent
The white paper also discusses the interplay between 
neutrinoless double-beta decay, experiments at the Large Hadron Collider and results from astrophysics and cosmology
in probing simplified models of lepton-number violation at the TeV scale, and the generation of the matter-antimatter asymmetry via leptogenesis.

\vspace{0.1 cm}

\noindent
This white paper is prepared for the topical groups TF11 (Theory of Neutrino Physics), TF05 (Lattice Gauge Theory), RF04 (Baryon and Lepton Number Violating Processes), NF03 (Beyond the Standard Model) and NF05 (Neutrino Properties) 
within the Theory Frontier, Rare Processes and Precision Frontier, and
Neutrino Physics Frontier of the U.S. Community Study
on the Future of Particle Physics (Snowmass 2021).

\end{abstract}
\pacs{}

\maketitle

\tableofcontents

\newpage
\section*{Executive Summary}
\label{sec:exec}
\noindent
\noindent
The next generation of tonne-scale neutrinoless double-beta ($0\nu\beta\beta$) decay experiments has the opportunity to answer fundamental questions about the nature of neutrino masses, with profound implications for our
understanding of the mechanism by which neutrino mass is generated and of the origin of the matter-antimatter asymmetry in the Universe. 
While the observation of $0\nu\beta\beta$ decay
will imply that neutrinos are Majorana particles, $0\nu\beta\beta$ experiments are sensitive to a variety of lepton-number-violating (LNV) mechanisms, including the standard scenario driven by the exchange of light Majorana neutrinos, low-scale seesaw scenarios with light sterile neutrinos below the electroweak scale, and models of physics beyond the Standard Model (BSM) with new degrees of freedom at the TeV scale. The observation of $0\nu\beta\beta$ decay could even help solve the flavor puzzle because different classes of flavor models predict different rates of decay.

The interpretation of 
$0\nu\beta\beta$ experiments and, in case of an observation, the 
solution of the ``inverse problem'' of identifying the microscopic mechanism behind a signal demand an ambitious theoretical program to: $a)$ further develop particle-physics models of LNV, including simplified models that go beyond the Majorana neutrino-mass paradigm, 
and test them against the results of current and future $0 \nu \beta \beta$ 
experiments, the Large Hadron Collider (LHC), and astrophysics and cosmology;
$b)$  compute $0\nu\beta\beta$ rates with minimal model dependence and quantifiable theoretical uncertainties by advancing progress in particle and nuclear effective field theories (EFTs), lattice quantum chromodynamics (QCD), and nuclear few- and many-body \textit{ab initio} methods. This white paper identifies three avenues that will allow an accurate matching of theory to experiment:
\begin{itemize}[leftmargin=*]
\item[$\circ$]{\emph{Bridging particle and nuclear physics with EFTs.} 
Simplified models of LNV at the TeV scale provide useful schemes for investigating correlated signals of LNV across multiple types of experiments, spanning wide energy and distance scales. For certain sources of  LNV, LHC experiments have the potential to surpass constraints from $0\nu\beta\beta$ experiments and  to explore complementary regions of parameter space. We encourage the LHC collaborations  to investigate simplified models for TeV-scale LNV and to make projections for the high-luminosity LHC (HL-LHC). LHC searches will help exclude or discover alternative TeV-scale interpretations of a $0\nu\beta\beta$ signal, and, through discovery, 
help falsify high-scale models of leptogenesis. Finally, both full and simplified BSM models can be matched to hadronic EFTs, which allow for a systematic expansion of the $0\nu\beta\beta$ rates.}
     
\item[$\circ$]{\emph{Lattice-QCD input for $0\nu\beta\beta$ decay.} The EFTs that systematically classify LNV in the few-nucleon sector need to be complemented with values for low-energy constants (LECs).  The lack of experimental input makes lattice QCD, which computes the relevant matrix elements directly, the only way to determine these LECs.  To systematically and reliably match lattice QCD matrix elements to EFT LECs, we require a) a targeted computational campaign, powered by high-performance computing, to obtain precise two-nucleon spectroscopy and matrix elements, and b) theoretical developments to clarify the path from lattice-QCD output to physical matrix elements in the two- and higher-nucleon sectors, within various high-scale models of LNV.}

\item[$\circ$]{\emph{Nuclear structure and the computation of $0\nu\beta\beta$ matrix elements.} With the systematic EFTs constrained by lattice QCD, uncertainties in nuclear matrix elements (NMEs) of relevance to experimental isotopes can be realistically assessed and reduced. 
Under the auspices of the Topical Collaboration on Nuclear Theory for Double-Beta Decay and Fundamental Symmetries, researchers began to shift from phenomenological approaches to \textit{ab initio} methods for the calculation of $0\nu\beta\beta$ NMEs, with the first wave of \textit{ab initio} results for $^{48}$Ca, $^{76}$Ge,
and $^{82}$Se appearing in the last few years.
Work is already underway to validate and improve the approximations used in these calculations. The computation of next-generation NMEs for $0\nu\beta\beta$ candidate nuclei will require considerable amounts of computing time as well as investments in the
development of many-body codes to ensure that the allocated time is used efficiently. Parallel advances in nuclear EFTs and lattice QCD are required to guarantee that the nuclear interactions and transition operators are constructed at the same order and in the same regularization scheme. Finally, the community needs to address deep questions about the 
implementation of EFT interactions and operators in traditional many-body methods.

}     
\end{itemize}

\newpage

\section{Introduction}
\noindent
The observation of neutrino
oscillations around the turn of the last century
\cite{Super-Kamiokande:1998kpq, SNO:2001kpb,KamLAND:2002uet} 
constituted the first direct laboratory evidence of BSM physics. 
The present and future generation of neutrino oscillation experiments, including the Deep Underground Neutrino Experiment (DUNE), the flagship U.S.  experiment in high-energy physics, and the short-baseline neutrino (SBN) program at Fermilab, aim at answering fundamental questions on the properties of massive neutrinos, such as the neutrino mass ordering and the amount of violation of the symmetry of charge conjugation and parity (CP) in the lepton sector. The era of precision neutrino physics will start to test the three-neutrino paradigm and test hints of additional light states~\cite{MiniBooNE:2018esg,Barinov:2021asz,MicroBooNE:2021zai}.
Neutrino oscillation experiments alone,
however, cannot answer one of the most important questions on the nature of massive neutrinos: are neutrinos Dirac particles or Majorana particles?
The answer to this question carries profound implications for other open problems in the SM, from the origin of the matter-antimatter asymmetry in leptogenesis scenarios~\cite{Davidson:2008bu} to identifying the energy scale of BSM physics~\cite{Weinberg:1979sa}. 

If neutrinos are Majorana particles, lepton number ($L$) is not a good symmetry of perturbation theory
and can be violated in processes such as $0\nu\beta\beta$ decay~\cite{Furry:1939qr}, in which two neutrons inside a nucleus turn into two protons with the emission of two electrons and zero neutrinos, violating $L$ by two units. 
$0\nu\beta\beta$ experiments are the only feasible way to observe LNV directly induced by light Majorana neutrinos. For this reason,
a vigorous experimental program is in place to look for $0\nu\beta\beta$ decay in a variety of nuclear isotopes. The present generation of experiments has already obtained impressive limits on the decay's half-life~\cite{NEMO-3:2015jgm,KamLAND-Zen:2016pfg,EXO-200:2019rkq,Majorana:2019nbd,GERDA:2020xhi,CUORE:2021gpk,CUPID:2020aow,KamLAND-Zen:2022tow}. The goal of the next generation of experiments is to improve the limits by a factor of 100 in order to reach half-lives in the range of $10^{28}$ yr,
which, in the light-Majorana-neutrino exchange scenario, is equivalent to probing the combination $|m_{\beta\beta}| = |\sum m_i U_{ei}^2|$ down to about $0.01$ eV and covering the entire inverted hierarchy region. Here, $|m_{\beta\beta}|$ is called the effective Majorana neutrino mass, $m_i$ is the mass of the neutrino mass eigenstate $i$ and $U_{ei}$ are the elements of the Pontecorvo-Maki-Nakagawa-Sato (PMNS)
matrix. 
In addition to $0\nu\beta\beta$ decay, LNV searches at the LHC
 and other high-energy colliders~\cite{ATLAS:2018dcj,CMS:2018jxx,ATLAS:2019kpx,ATLAS:2020wop}
 can probe competing LNV mechanisms and suggest the Majorana nature of neutrinos. 
Searches for LNV and $0\nu\beta\beta$ decay thus complement the neutrino oscillation program, and are necessary to obtain a complete picture of neutrino physics.  

A detection of $0\nu\beta\beta$ decay  raises further questions. Foremost is the ``inverse problem":
\begin{itemize} 
\item[$i)$] Are Majorana neutrino masses the correct physical explanation for such a detection? If so, 
what are the implications of such a detection for theoretical models of neutrino masses?
If not, what are alternative interpretations? How can they be excluded?
\end{itemize}
While the observation of $0\nu\beta\beta$ decay, as guaranteed by the Schechter-Valle ``black-box'' theorem~\cite{Schechter:1981bd}, would imply that neutrinos are Majorana particles, 
the minimum size of the Majorana mass consistent with such an observation is extremely tiny and entirely irrelevant to the question of whether Majorana neutrino masses are the explanation for an observed 
$0\nu\beta\beta$ decay.\footnote{The shift in the Majorana neutrino mass, derived using the current Kamland-ZEN experimental limit $T^{0 \nu}_{1/2}>1.07\times 10^{26}$ year~\cite{KamLAND-Zen:2016pfg}, is roughly only $2\times 10^{-28}$~eV~\cite{Duerr:2011zd}.}

Moreover, 
the robust interpretation of $0\nu\beta\beta$ experiments in terms of microscopic parameters in the neutrino sector 
and their connection to oscillation experiments depends on: 
\begin{itemize} 
\item[$ii)$] The theoretical assumptions on the dominant contribution to the $0\nu\beta\beta$ rate;
\item[$iii)$] The  uncertainties involved in determining LNV interactions of hadrons and leptons in terms of quark-level couplings;
\item[$iv)$] And, 
the uncertainties in the calculation of NMEs of $0\nu\beta\beta$ transition operators. 
\end{itemize}
Understanding the implications of $0\nu\beta\beta$ experiments on high-energy particle physics thus requires
an ambitious theoretical program on several fronts involving: a) particle and nuclear EFTs, lattice QCD, and nuclear few- and many-body \textit{ab initio} methods
to deliver $0\nu\beta\beta$ rates with minimal model dependence and quantifiable theoretical uncertainties; and b) further development of particle-physics models of LNV, including simplified models, that go beyond the Majorana neutrino mass paradigm, 
and confront them against current and future measurements in $0 \nu \beta \beta$ 
experiments, the LHC, and results from astrophysics and cosmology, in order to make headway on the ``inverse problem''. In the context of the progress made to date in each of these theoretical frontiers, the milestones to be achieved in the near future will be identified in this white paper.

\section{Bridging Particle and Nuclear Physics with EFTs}\label{bridge}
\noindent
We start with an overview  of the implications of $0\nu\beta\beta$ experiments on particle physics and cosmology.
Section~\ref{sec:numass} discusses the constraints of $0\nu\beta\beta$
on neutrino mass models, including the type I, II and III seesaw scenarios and flavor models built to explain the structure of the PMNS matrix. In Section~\ref{sec:TeV}, we extend the discussion to scenarios with LNV at the TeV scale, and examine the interplay between  $0\nu\beta\beta$ decay and detector experiments at high-energy hadron colliders, including ATLAS and CMS, and new efforts designed to look for long-lived particles. Section \ref{sec:cosmo} discusses the complementarity between cosmology and $0\nu\beta\beta$ decay and the impact of LNV at the TeV  
scale on the generation of the matter-antimatter asymmetry via leptogenesis. Finally, in Section \ref{sec:EFTs-for-LNV} we review recent progress on how to systematically connect LNV models with $0\nu\beta\beta$ experiments using a tower of EFTs. 

\subsection{Neutrinoless Double-Beta Decay and Neutrino  Model Building}\label{sec:numass}
\begin{figure}
    \centering
    \includegraphics[width=\textwidth]{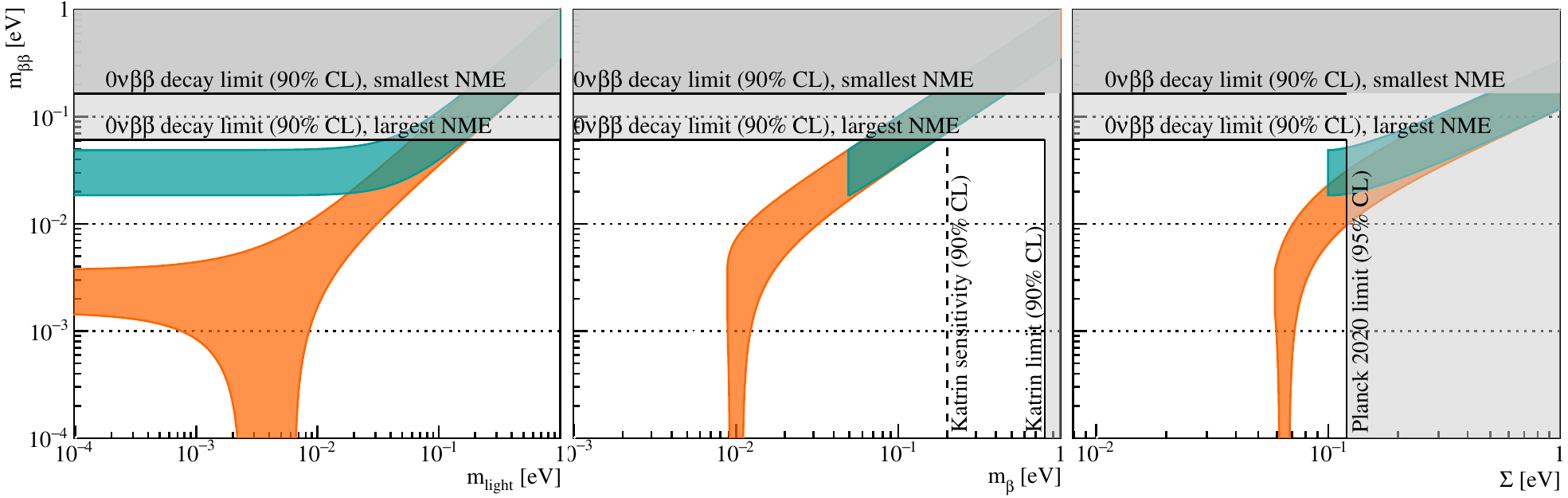}
    \caption{
    Electron-neutrino effective Majorana mass $|m_{\beta\beta}|$ as a function of the mass of the lightest neutrino $m_{\rm light}$ (left panel), of the combination of neutrino masses and mixing  probed in $\beta$ decays, $m_{\beta}$ (middle panel), and of sum of neutrino masses $\Sigma$, probed in cosmology. The blue and orange bands 
    denote 
    the inverted and normal orderings, respectively, and are obtained by marginalizing over the Majorana phases, while fixing the other parameters in the PMNS matrix to their central values. The gray horizontal bands denote the current $0\nu\beta\beta$ limits, for two choices of NMEs. The solid and dashed vertical lines in the middle panel are the 
    present and projected limits on $m_{\beta}$ from the KATRIN experiment~\cite{Aker:2021gma}, while the gray vertical band in the right panel denotes the limit from Planck~\cite{Planck:2018vyg}. 
    The figure was taken from Ref.~\cite{Agostini:2022zub}.}
    \label{fig:masses}
\end{figure}

Neutrino masses are direct evidence of BSM physics, and understanding their origin may lead to further discoveries. 
In the minimal scenario of three light Majorana neutrinos, the neutrino sector is described by three masses and a $3\times3$ unitary matrix, the PMNS matrix, which is parameterized by three mixing angles, one Dirac CP phase, and two Majorana CP phases. 
Oscillation experiments have accurately measured two mass differences and the three mixing angles~\cite{ParticleDataGroup:2020ssz}. DUNE will pin down the Dirac phase and the neutrino mass ordering. 

The absolute scale of neutrino masses can be probed by $0\nu\beta\beta$ experiments, by precise reconstruction of the electron kinematics in $\beta$ decays, and by the imprint of neutrinos on the evolution of the Universe.  
The latter suite of consequences are of course predicated on the theoretical assumption that neutrinos are stable on the timescale of the cosmological epoch in question. 
These three set of probes are sensitive to three different combinations of neutrino masses and PMNS matrix elements, 
$|m_{\beta\beta}| = |\sum_i U_{ei}^2 m_i|$, $m_{\beta} = \sqrt{\sum_i |U_{ei}|^2 m^2_i}$ and $\Sigma = \sum_i m_i$, respectively. 
Figure~\ref{fig:masses}, taken from a recent $0\nu\beta\beta$ review~\cite{Agostini:2022zub},
summarizes our current knowledge of these three combinations using an up-to-date extraction of $U_{ei}$~\cite{ParticleDataGroup:2020ssz}.
The bands are obtained by assuming the central values of the neutrino oscillation parameters and marginalizing over the unknown Majorana phases. The blue and orange bands correspond to the inverted and normal orderings, respectively. 
The gray horizontal bands denote current limits from $0\nu\beta\beta$ experiments, currently led by KamLAND-Zen~\cite{KamLAND-Zen:2016pfg}, with different assumptions on the NMEs.
The more recent KamLAND-Zen result,
$T^{0 \nu}_{1/2}>2.3\times 10^{26}$ at 90\% C.L.~\cite{KamLAND-Zen:2022tow}, would strengthen the constraints by a factor of 1.5.
The vertical gray regions are excluded by the KATRIN experiment (middle panel)~\cite{Aker:2021gma} and by Planck space observatory (right panel)~\cite{Planck:2018vyg}.

Figure~\ref{fig:masses} illustrates the complementarity of experiments that probe neutrino masses. In the next generation, tonne-scale $0\nu\beta\beta$ experiments will reach $|m_{\beta\beta}| \sim 0.015$ eV, possibly covering the entire inverted-ordering region
(with some uncertainty due to the NMEs, as will be discussed later).
The dashed line on the middle panel shows the projected sensitivity of the full KATRIN dataset, $m_\beta \sim 0.2$ eV. 
The ECHo experiment, based on electron capture on $^{163}$Ho, will also access the sub-eV region~\cite{Gastaldo:2017edk}. Even more exciting is the prospect of reaching $m_{\beta} \sim 0.04$ eV in the tritium $\beta$ decay experiment within Project 8~\cite{Project8:2017nal},
and of probing masses down to $0.01$ eV in the Cosmic Neutrino Background experiment PTOLEMY~\cite{PTOLEMY:2019hkd}, covering the entire parameter space. This is therefore a very exciting time in neutrino physics, with the concrete possibility to determine the absolute neutrino-mass scale!

These observations will also have an important impact on theoretical models that aim to provide a rationale behind the smallness of the neutrino masses and the observed pattern of neutrino mixing. 
A natural explanation of the smallness of neutrino masses comes from the popular seesaw mechanism~\cite{Minkowski:1977sc,Ramond:1979py,Gell-Mann:1979vob,Yanagida:1979as,Mohapatra:1979ia,Schechter:1980gr}, which predict that neutrinos are Majorana fermions.
Therefore an observation of $0\nu\beta\beta$ decay will be fundamental in our  quest for the neutrino-mass mechanism. In fact, minimal
 realizations of the seesaw mechanism introduce extra degrees of freedom 
 which could leave an imprint in $0\nu\beta\beta$ experiments~\cite{Blennow:2010th,Mitra:2011qr}.
In the following we will focus on minimal realizations of the seesaw mechanism, see Sec.~\ref{sec:TeV} for other sources of LNV that can be probed with $0\nu\beta\beta$ decay.
 
 In a minimal type I seesaw scenario~\cite{Minkowski:1977sc,Yanagida:1979as,Mohapatra:1979ia,Gell-Mann:1979vob}, additional sterile neutrinos are introduced.  Two regimes can be distinguished depending on whether the sterile mass $m_N$ is larger or smaller than the average momentum exchange squared of the process $\langle p^2\rangle\sim (100~\text{MeV})^2$~\cite{Blennow:2010th,Bolton:2019pcu}. 
 If $m_N^2>\langle p^2\rangle$,
 the contribution from a ‘heavy’ sterile neutrino is suppressed
by $1/m_N$ and by the active-sterile mixing matrix-element squared. If $m_N^2\gg\langle p^2\rangle$, the heavy sterile neutrinos are integrated out and the contribution from the light active neutrinos thus dominate~\cite{Bolton:2019pcu,Blennow:2010th}. If $m_N^2\ll\langle p^2\rangle$, the sterile neutrino contributes to $0\nu\beta\beta$ decay like an active neutrino and the $0\nu\beta\beta$ rate is suppressed, similar to the GIM suppression
in flavor-violating processes~\cite{Blennow:2010th}. Finally, introducing multiple sterile neutrinos, some with mass above and some below $\langle p^2\rangle$, which could be, for example, realized in an extended seesaw scenario~\cite{Kang:2006sn}, leads to a rich phenomenology. This is because the `light’ sterile-neutrino contribution may even dominate over the light active-neutrino
contribution and lead to rates for $0\nu\beta\beta$ decay larger than expected from active neutrinos only~\cite{Lopez-Pavon:2012yda,Jha:2021oxl,Abada:2018qok,Bolton:2019pcu,Deppisch:2020ztt}.

In type II seesaw models~\cite{Magg:1980ut,Schechter:1980gr,Wetterich:1981bx,Lazarides:1980nt,Mohapatra:1980yp},  a scalar
$SU(2)_L$ triplet with hypercharge 2 is introduced. However, the contribution of the  scalar to  the $0\nu\beta\beta$ rate is suppressed with respect
to the neutrino one such that, in this scenario, the light active-neutrino contribution dominates
and the usual description of $0\nu\beta\beta$ decay applies~\cite{Blennow:2010th}.
In type III seesaw models~\cite{Foot:1988aq,Ma:1998dn,Ma:2002pf,Hambye:2003rt}, the Standard Model (SM) is extended
by a  $SU(2)_L$ fermion triplet with  zero hypercharge. 
The $0\nu\beta\beta$ phenomenology of the type III seesaw is  completely analogous to
that of the type I seesaw, with the neutral component of the triplet playing the role of the sterile neutrino. However, since the triplet also has charged components, stringent lower
bounds on its mass exist and in practice only the heavy mass eigenstate regime is available~\cite{Blennow:2010th,ATLAS:2022yhd}.

Depending on the results of $0\nu\beta\beta$ experiments, 
coupled with advances in neutrino oscillation experiments and cosmology,
there are  different implications for  minimal seesaw mechanisms~\cite{Blennow:2010th}. These will also help to answer the ``inverse problem":
\begin{itemize}[leftmargin=*]
\item[$\circ$]{\emph{The $0\nu\beta\beta$ process is observed to be in agreement with the predicted rates. }}
This means that the 
 light active neutrinos dominate the $0\nu\beta\beta$ rate. This also implies that   new physics introduced in the minimal seesaw models is above the nuclear scale such  
that its contribution is suppressed.
\item[$\circ$]{\emph{The $0\nu\beta\beta$ process is observed to be smaller than the predicted rates.}} 
This indicates that there is a partial cancellation between the active and sterile neutrino
contributions, which can be achieved if the sterile masses in minimal seesaw scenarios are around the nuclear scale.
\item[$\circ$]{\emph{The $0\nu\beta\beta$ process is observed to be larger than the predicted rates.}} In this
situation the light active neutrinos cannot dominate the $0\nu\beta\beta$ rate. In minimal type I seesaw scenarios, sterile
neutrinos lighter than or around the nuclear scale are required in this case. To avoid a cancellation between the contribution from active neutrinos and from sterile neutrinos,
a cancellation between sterile
neutrinos both above and below the nuclear scale  or between the
extra neutrinos and a type II or III seesaw contribution needs to be achieved.
\item[$\circ$]{\emph{The $0\nu\beta\beta$  process is not observed but was predicted.}} This could imply that neutrinos are Dirac particles or that neutrinos are Majorana particles but there are sterile neutrinos  below the nuclear scale. In this case there are
GIM-like cancellations  and the $0\nu\beta\beta$ rate becomes unobservably small.
\item[$\circ$]{\emph{The $0\nu\beta\beta$ process is not observed at next generation experiments and was predicted to  be unobservably small.}} In this case we cannot draw any conclusion about neutrino masses. Neutrinos could be Dirac particles, or they are Majorana fermions but  the $0\nu\beta\beta$ rate suffers from cancellations, potentially even involving new-physics contributions.
\end{itemize}
Some of these statements can already be tested with the next generation of $0\nu\beta\beta$ experiments which will probe the region predicted for inverted ordering.
Apart from these minimal seesaw mechanisms, other models
which give rise to LNV 
can be probed with $0\nu\beta\beta$ decay, see Sec.~\ref{sec:TeV}.
It is also important to stress that predictions for $0\nu\beta\beta$ decay can vary considerably in non-minimal models~\cite{Tello:2010am}.

\begin{figure}
    \centering
    \includegraphics[scale=0.55]{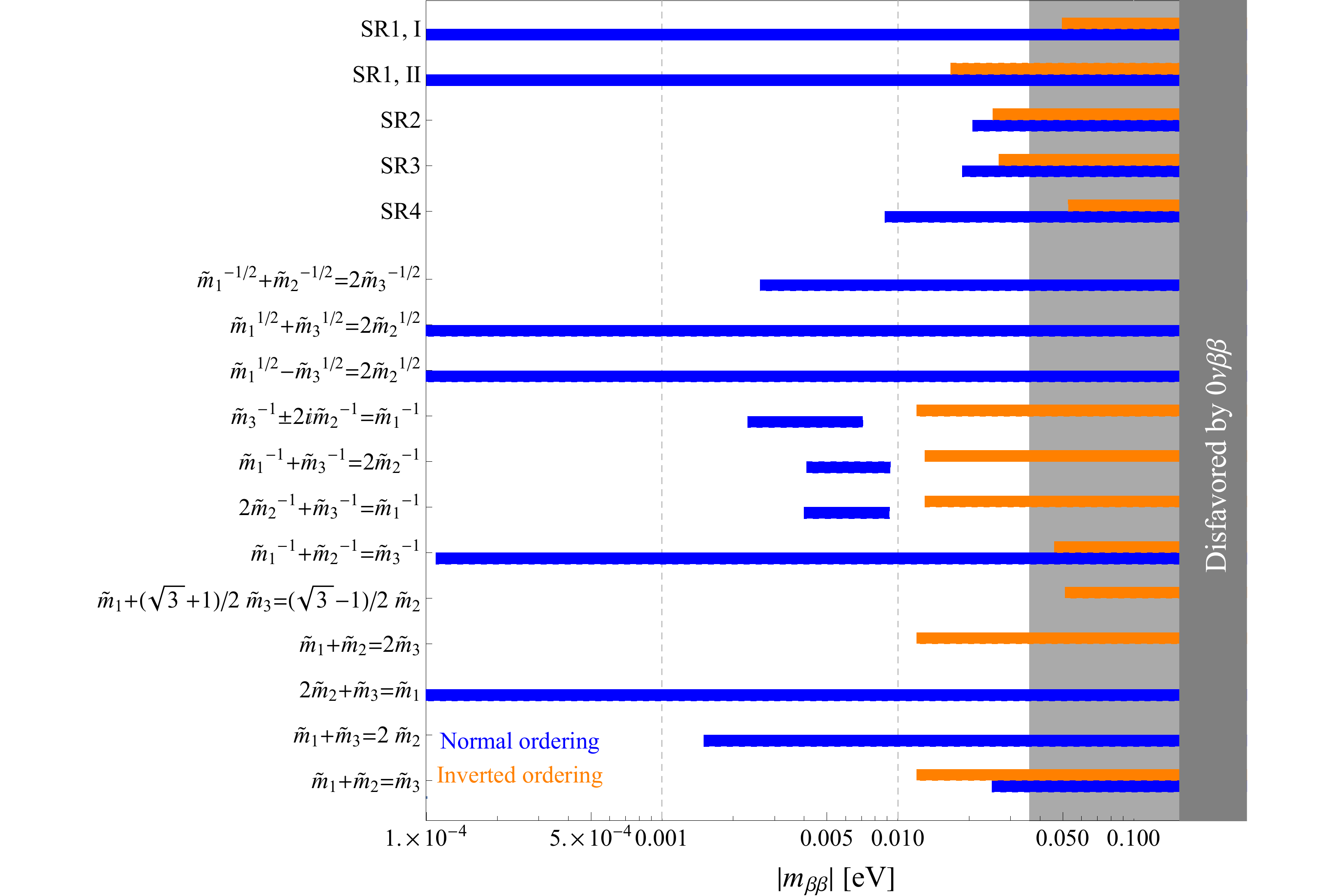}
    \caption{Representation of the predictions for $|m_{\beta\beta}|$ from different mass sum rules. The upper five mass sum rules (SR) have been derived in models based on modular symmetries~\cite{Gehrlein:2020jnr}, the lower twelve mass sum rules come from models based on discrete symmetries~\cite{King:2013psa}. The  grey regions show the constraints on $|m_{\beta\beta}|$ from Ref.~\cite{KamLAND-Zen:2022tow}  using commonly-adopted nuclear matrix-element calculations.
    }
    \label{fig:flavormodels}
\end{figure}
In addition to the open question of the origin of the neutrino masses, the rationale behind the observed neutrino-mixing pattern remains to be found.
Some flavor models based on discrete symmetries provide
theoretical predictions for the absolute scale of neutrino masses  as well as for $|m_{\beta\beta}|$.  Of particular interest 
are models  that predict a correlation between the three complex neutrino mass
eigenvalues, $\tilde{m}_i=m_i \, \text{exp}(\text{i}\alpha_i)$ 
(with Majorana phases
$\alpha_1$, $\alpha_2$ and with $\alpha_3$ chosen to be unphysical), e.g.,
$\tilde{m}_1+\tilde{m}_2+\tilde{m}_3=0$ or  
$\tilde{m}_1^{-1}+\tilde{m}_2^{-1}+\tilde{m}_3^{-1}=0$. These relations appear due to a parameter reduction when the three mass eigenvalues can be described by two model parameters only~\cite{Gehrlein:2017ryu}.
As these predictions involve the Majorana phases, these models can be studied best in $0\nu\beta\beta$ experiments. Additionally, these sum rules also lead to a lower bound on the lightest mass. Furthermore, some sum rules  even predict the neutrino-mass ordering. 

Neutrino-mass sum rules are present in over 60 flavor models in the literature, hence
providing a motivated target for experiments. 
Studies of their predictions have been conducted in Refs.~\cite{Bazzocchi:2009da, Altarelli:2009kr, Barry:2010zk, Chen:2009um, Altarelli:2008bg, Hirsch:2008rp}
but only in recent years more systematic categorisations
and analyses  have been  
performed~\cite{Barry:2010yk,Dorame:2011eb,King:2013psa, Gehrlein:2016wlc, Gehrlein:2015ena}.
Up to now 12 different mass sum rules have been identified that have different predictions in the $|m_{\beta\beta}|-m_{light}$ plane, allowing the ability to probe classes of flavor models as well as to distinguish them from each other with measurements of these observables.
In Ref.~\cite{Gehrlein:2020jnr}, five new mass sum rules have been identified in  
flavor models based on modular symmetries~\cite{Feruglio:2017spp}, which also involve mixing parameters, thus enhancing their testability at a variety of neutrino experiments.
Certain classes of mass sum rules predict a value of the lightest neutrino mass close to the current upper limit from cosmology and can, therefore, be probed in the near future. 
Similarly, whole groups of neutrino flavor models can be tested when $0\nu\beta\beta$ experiments reach a certain sensitivity to $|m_{\beta\beta}|$~\cite{Agostini:2015dna}, see Fig.~\ref{fig:flavormodels}.
These predictions can, therefore, be used to plan stages of experiments. 
Finally, among flavor models that can be probed with $0\nu\beta\beta$ experiments are those that predict definite  values for the Majorana phases such that they are 
either CP conserving or are in line with predictive schemes combining flavor and generalized CP
symmetries~\cite{Pascoli:2007qh,Penedo:2018kpc}.

\subsection{Lepton Number Violation at the TeV Scale}\label{sec:TeV}
It has been known for some time that $0\nu\beta \beta$ experiments place significant constraints on particle physics models that have LNV. 
If the characteristic mass scale $\Lambda$ of LNV is above the electroweak scale, then, at energies much lower compared to $\Lambda$, the effects on LNV can be described as a series of higher dimensional operators invariant under the SM gauge symmetries, suppressed by the scale $\Lambda$. If the LNV operators additionally conserve baryon number,\footnote{This article does not consider the other logical possibility.} then they have an odd operator dimension~\cite{Kobach:2016ami}, as discussed further in Sec.~\ref{sec:EFTs-for-LNV}. After the Weinberg operator that is of dimension-5, the next set of LNV operators occurs successively at dimension-7, dimension-9, and so on. 

\begin{figure}[!t] 
\centering
\includegraphics[width=0.575\linewidth]{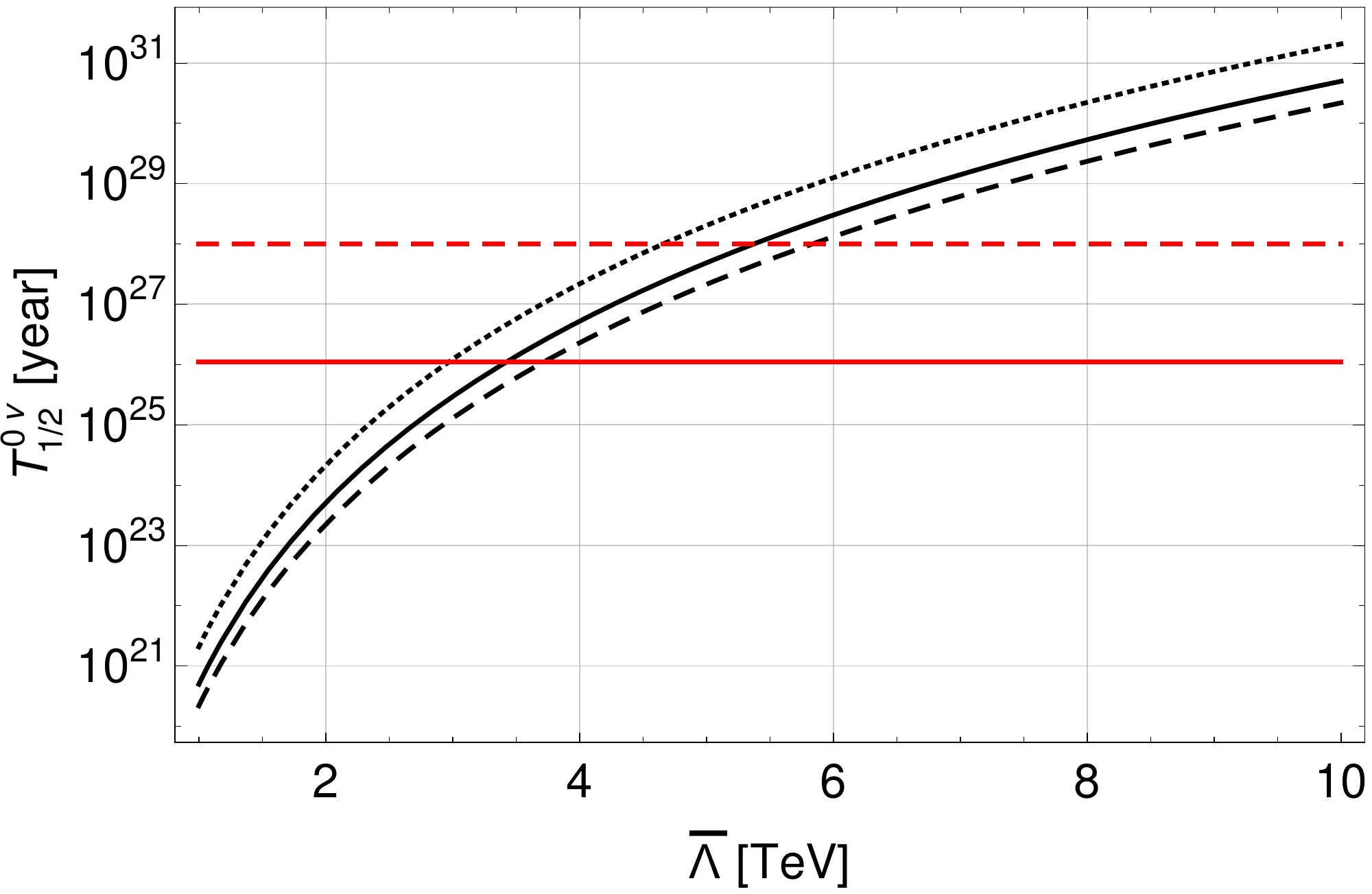}	
\caption{ $0\nu\beta\beta$
half-life, $T_{1/2}^{0\nu}$,
as a function of an effective LNV scale $\bar{\Lambda}$, 
with the assumption that the rate is dominated by a particular dimension-9 LNV operator.  
The solid, dashed, and dotted black curves are obtained assuming different $O(1)$ values for 
certain unknown hadronic LECs. The solid red and dashed red lines correspond to the current and projected constraints on the half-life from the KamLAND-Zen experiment~\cite{KamLAND-Zen:2016pfg}, 
$T_{1/2}^{0\nu}>1.07\times 10^{26}~\text{year}$ for $^{136}$Xe, 
and future tonne-scale experiments~\cite{Kharusi:2018eqi,Abgrall:2017syy,CUPIDInterestGroup:2019inu,Paton:2019kgy,Chen:2016qcd,Adams:2020cye} 
that aim to improve the half-life sensitivity by about 2 orders of magnitude, reaching  $T_{1/2}^{0\nu}> 10^{28}~\text{year}$.
Figure is taken from Ref.~\cite{Graesser:2022nkv}. 
} 
\label{fig:DBD-sensitivity}
\end{figure}

In the case of dimension-9 LNV operators, current and future limits from $0\nu\beta \beta$ experiments generically probe the multi-TeV scale~\cite{Pas:2000vn,Prezeau:2003xn,deGouvea:2007qla, Cirigliano:2018yza}, as illustrated in Fig.~\ref{fig:DBD-sensitivity}
for a specific dimension-9 operator. 
In the case of dimension-7 operators, limits range from $\sim $ 40 TeV up to hundreds of TeV~\cite{Cirigliano:2017djv}. 
While some of these scales are significantly larger than the TeV scale, they come with several caveats: a) in quoting limits, the Wilson coefficient of the LNV operator has been normalized to one; and, b) 
the theoretical extraction of limits on $\Lambda$ arising from $0 \nu \beta \beta$ half-lives comes with considerable hadronic uncertainties, as is evident from Fig.~\ref{fig:DBD-sensitivity}. These uncertainties are discussed in the Sections that follow. Because of the first caveat, the physical masses of the particles that generate these operators can be much lower than the quoted bound on $\Lambda$ whenever small coupling constants are involved.

Broadly speaking, $0\nu\beta \beta$ experiments probe a scale of LNV BSM physics that can also be probed by the LHC and future colliders, in ways that are complementary to each other~\cite{Cai:2017mow}. Consequently, discoveries or exclusions from the LHC on LNV interactions provide important insight on one aspect of the ``inverse problem,'' namely whether an observation of $0\nu\beta \beta$ decay can 
be explained by physics  at the TeV-scale.

Signals of LNV at the LHC can be organized according to whether or not the exotic particles mediating LNV can be produced on-shell at LHC energies:
\begin{itemize}[leftmargin=*]
\item[$\circ$] If not, the physics of LNV at the LHC is rather simple, since it can be described in a  
model-independent way by $SU(2)_L\times U(1)_Y$-invariant LNV operators.
At dimension-5, the Weinberg operator manifests as same-sign $W^\pm W^\pm$ scattering into same-sign leptons, i.e., $W^\pm W^\pm \to \ell^\pm_i \ell^\pm_j$~\cite{Fuks:2020zbm}.
With the full Run II data set, the CMS experiment has set an upper bound on the quantity $\vert m_{\mu\mu}\vert=\vert \sum_i U_{\mu i}^2 m_i\vert$  of about 11 GeV at 95\% CL~\cite{CMS:2022zsu}.  In the case of dimension-7 operators, 
a sizable fraction of the operators do not involve quarks, so many cannot be directly accessed at the LHC.
A subset of dimension-7 that can be tested at the LHC correspond to the case of same-sign $W^\pm W^\pm$ scattering into same-sign leptons $\ell^\pm_i \ell^\pm_j$  when mediated by a far off-shell Majorana neutrino~\cite{Dicus:1991fk,Datta:1993nm,Aoki:2020til,Fuks:2020att}. With the full Run II data set, the CMS experiment exclude $m_4 = 10$ TeV Majorana neutrinos with with $\vert V_{\mu 4}\vert^2 > 0.1$ at 95\% CL~\cite{CMS:2022zsu}. Some of the operators do involve quarks, but given that $0\nu \beta \beta$ decay already 
constrains the scale $\Lambda$ to be 40 TeV-100s TeV, it is unlikely that the LHC can improve on these limits.

Dimension-9 operators can, in principle, be probed at the LHC, since many of these operators contain 
four quarks. Here, LNV signals include the well-known final states having two leptons of the same charge (``same-sign dilepton'' (SS)). 
However, the $SU(2)_L \times U(1)_Y$-invariance of the operator(s) may also imply signals in
single-lepton + missing energy (MET), and jets + MET final states, with the MET provided by escaping neutrinos~\cite{Graesser:2016bpz}. While the rate for these latter two processes depend only on a Clebsch-Gordan coefficient and, therefore, can be comparable to the SS dilepton signal, the relatively higher backgrounds in these channels may make detection challenging. Another challenge is that the signal rate for SS dilepton production, as well as these other channels, is suppressed by the phase space for $2 \rightarrow 4$ multiple-particle production. For this reason we 
expect that in this limit of ``out-of-reach'' mediators, LHC exclusion limits on the scale $\Lambda$ of LNV violation to be weaker than $0\nu \beta \beta$~\cite{Fuks:2020att,Fuks:2020zbm,CMS:2022zsu}. At the same time, collider experiments are sensitive to flavor channels that are otherwise inaccessible in $0\nu\beta\beta$ experiments.
\item[$\circ$] If the mediators responsible for LNV can be produced on-shell at the LHC, then the sensitivity of the LHC to LNV is dramatically improved, because the production cross section into LNV final states can be enhanced due to the $s-$channel production of new particles.
(This remains true even if only a subset of the mediating states are kinematically accessible at colliders~\cite{Ruiz:2017nip,Nemevsek:2018bbt}.)
The complementarity between LHC and $0\nu \beta \beta$ experiments is, however, model-dependent as it hinges on the specifics of the mass spectrum of the exotic particles and the nature of the LNV interactions. For examples and summaries, see Refs.~\cite{Bajc:2007zf,Tello:2010am,Peng:2015haa,Cai:2017mow,Graesser:2022nkv} and references therein.
\end{itemize}
The last point implies that the EFT framework based on higher-dimensional operators, which allows one to describe $0\nu\beta\beta$ decay in a semi-model-independent way, breaks down for $\sqrt{s}\sim\Lambda\sim $ TeV. 
On the other hand, disentangling all  possible LNV mechanisms using $0\nu\beta\beta$ observables alone would be a very challenging task~\cite{Cirigliano:2017djv}, making it important to consider LNV collider signatures alongside $0\nu\beta\beta$ decay. In this context, 
minimal scenarios of LNV can play a useful role in assessing the complementarity between collider probes and $0\nu\beta\beta$ decay. Well-motivated simplified models of neutrino masses and LNV allow one to unambiguously assess their effects at the colliders, 
while the EFT framework can still be used to determine the contributions to $0\nu\beta\beta$ decay after matching the model to the EFT. 
Such studies of simplified models not only allow one to determine the relative sensitivities that can be reached at high and low energies, but can also identify collider signatures that could help disentangle different scenarios from one another.

Given the lack of experimental or theoretical guidance, it is important to take a broad approach to the potential sources of LNV and consider a range of simplified LNV models. 
Numerous analyses combining  $0\nu\beta\beta$ and collider probes already exist~\cite{Tello:2010am,Nemevsek:2011aa,Tello:2012qda,Peng:2015haa,Deppisch:2015qwa,Lindner:2016lpp,Fuks:2020att,Fuks:2020zbm}, though the description of $0\nu\beta\beta$ decay can be updated. 
Interesting minimal scenarios of LNV that could be studied in this way include the phenomenological Type-I seesaw and its combination with Type-II and -III seesaw models~\cite{Minkowski:1977sc, Yanagida:1979as, GellMann:1980vs, Glashow:1979nm,Mohapatra:1979ia, Shrock:1980ct, Schechter:1980gr,Konetschny:1977bn, Cheng:1980qt,Lazarides:1980nt,Schechter:1980gr, Mohapatra:1980yp}, which extend the Standard Model by additional sterile fermions, scalar triplets, or triplet fermion fields, respectively. Models with somewhat increased complexity include the Zee model of neutrino masses~\cite{Graf:2020cbf}, R-parity violating supersymmetry~\cite{Hirsch:1995zi,Hirsch:1995ek,Bolton:2021hje}, LNV scenarios involving leptoquarks, which in some cases can address anomalies observed in B decays~\cite{Cai:2017wry,Dev:2020qet,Deppisch:2016qqd},
scenarios that can account for the discrepancy with the anomalous magnetic dipole moment of the muon~\cite{Cirigliano:2021peb},
or the minimal left-right symmetric model~\cite{Pati:1974yy,Mohapatra:1974hk,Mohapatra:1974gc,Senjanovic:1975rk,Senjanovic:1978ev}, which in the limit of heavy gauge bosons reduces to a Type I+II seesaw model. Connections between 
$0 \nu \beta \beta$ experiments and signals at the LHC in the left-right symmetric model have also been studied~\cite{Bajc:2007zf,Tello:2010am,Cirigliano:2017djv,Li:2020flq,Li:2022cuq}.

The current and projected sensitivities of LHC searches in comparison to $0\nu\beta\beta$ searches 
is in itself highly sensitive to the simplified model for TeV-scale LNV~\cite{Cai:2017mow}. This observation is illustrated with a brief summary of such a comparison performed for two simplified models~\cite{Peng:2015haa,Graesser:2022nkv}. 

\begin{figure}[!t] 
\centering
\includegraphics[width=0.55\linewidth]{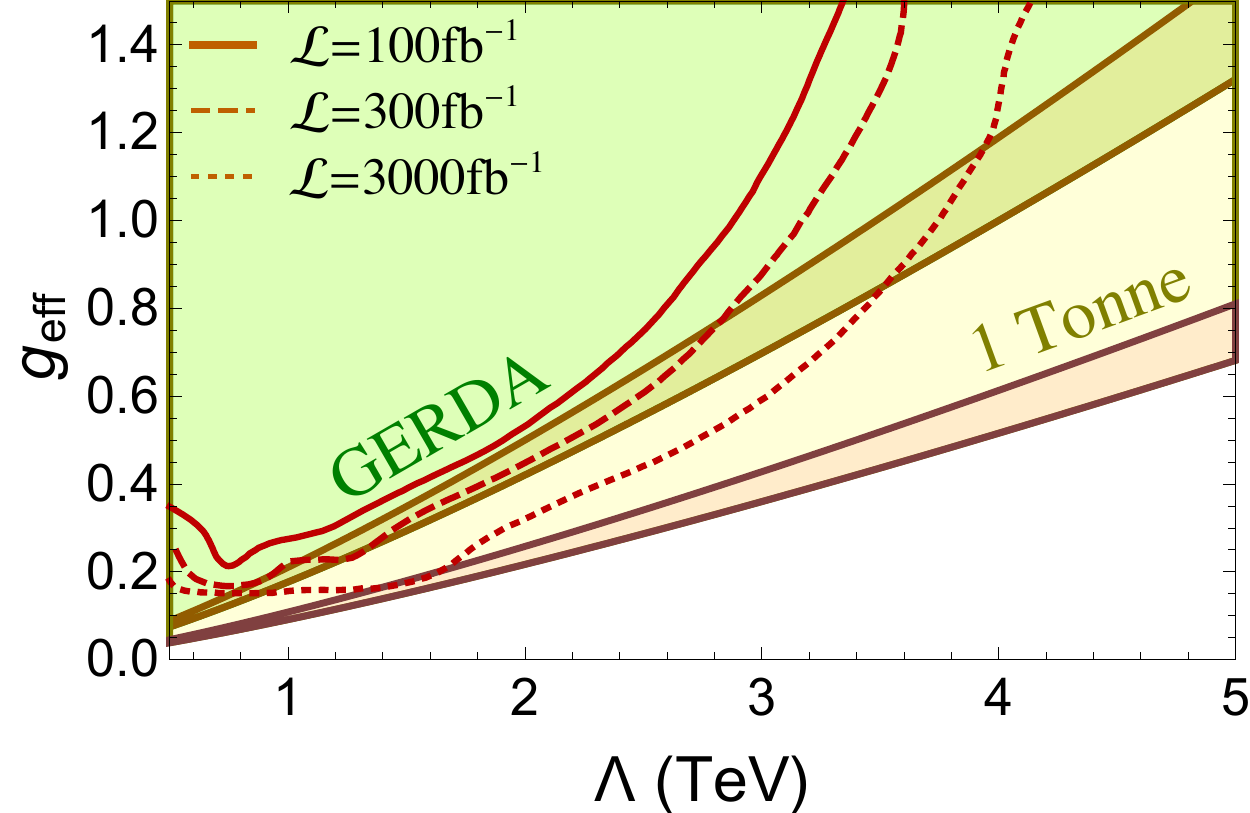}	
\caption{Past and future reach of $0\nu\beta \beta$ and LHC searches
for a R-parity violating low-energy supersymmetric inspired model, as functions of an effective
coupling $g_{\rm eff}$ and effective mass scale $\Lambda$. GERDA
exclusion~\cite{GERDA:2013vls} and 
prospective $^{76}$Ge tonne-scale sensitivity of
$T_{1/2} = 6 \times 10^{27}$ yr are indicated by upper and lower shaded regions, respectively.
Darker
shaded bands show the impact on $0\nu\beta \beta$ from varying a model mass parameter
by a factor of
two. LHC discovery reach for representative integrated luminosities are indicated by the solid, dashed, and dotted lines. Figure is taken from Ref.~\cite{Peng:2015haa}.}
\label{fig:Complementarity-plot1}
\end{figure}

Fig.~\ref{fig:Complementarity-plot1} shows the constraints from 
$0\nu \beta \beta$ decay and the LHC for a simplified model that is inspired by R-parity violating 
models of low-energy supersymmetry~\cite{Peng:2015haa}. Also shown are projections for future one-tonne $0\nu \beta \beta$ experiments and higher luminosity at the LHC. In this model, 
the reach
of the tonne-scale $0\nu \beta \beta$ decay experiments appears to out-perform
that of the HL-LHC. The LHC projections could be improved 
through better estimates of jet-fake and charge-flip
backgrounds.

\begin{figure}[!htb] 
\centering
\includegraphics[width=0.4\linewidth]{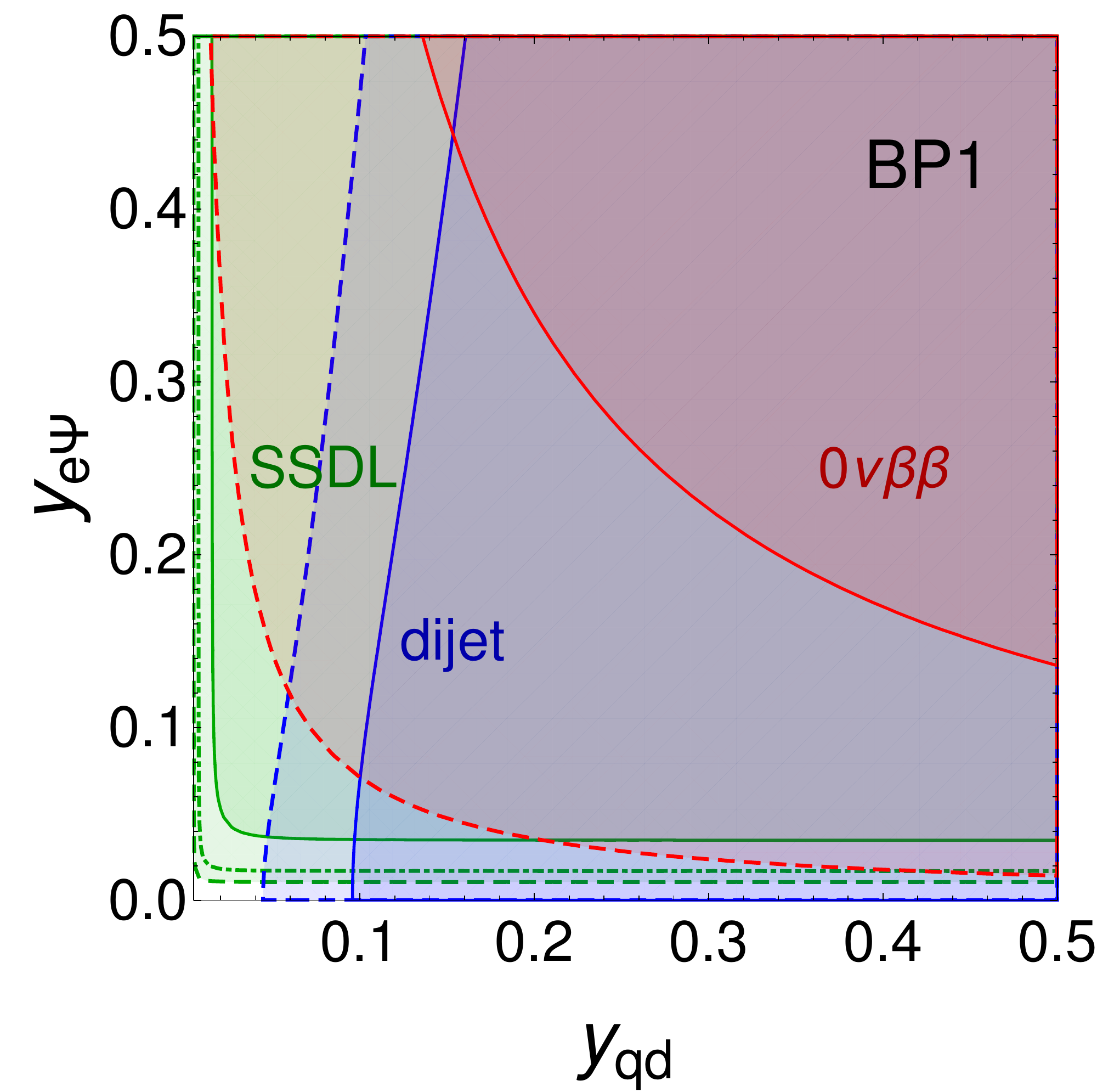}
\includegraphics[width=0.4\linewidth]{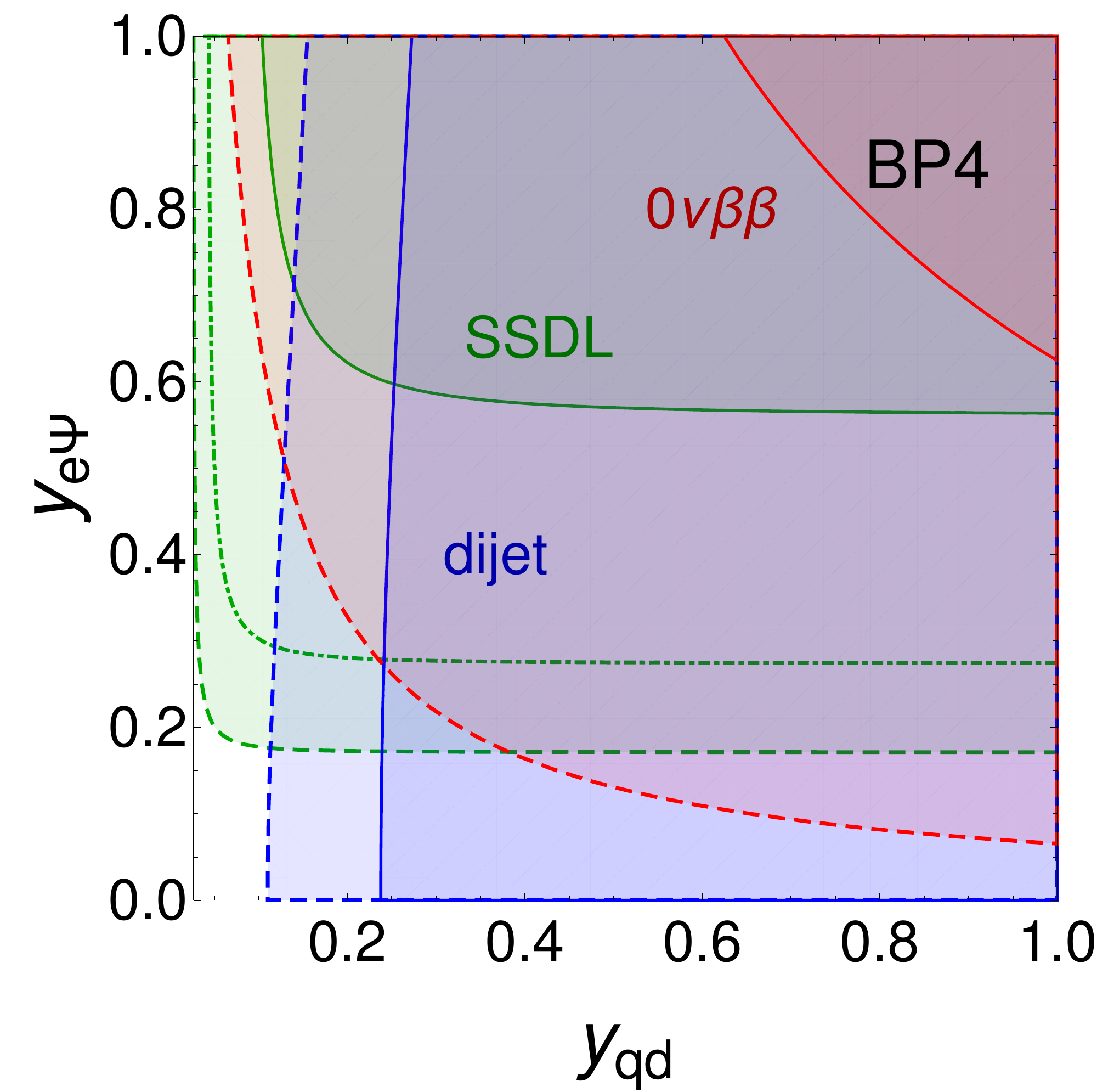}
\caption{The current and projected sensitivities for certain couplings in a leptoquark model with 
new particle masses at the TeV scale. The two panels correspond to two different choices for the mass spectra. 
The red regions give the current exclusion by the KamLAND-Zen experiment (solid curve) and projected exclusion at tonne-scale experiments (dashed curve). 
The blue regions give 
the current exclusion from dijet searches (solid curve) and 
projected exclusions with $3$ ab$^{-1}$ (dashed curve). The green regions give the current exclusion from same-sign 
dilepton (SSDL) searches (solid curve) and projected 
exclusions with $3$ ab$^{-1}$ (dashed curve). The green dot-dashed curves correspond to the $5\sigma$ discovery potential at the HL-LHC in the SSDL channel. Projections for the HL-LHC are based on the na\"ive assumption that signal-to-background remains unchanged from current searches. 
Figure is taken from Ref.~\cite{Graesser:2022nkv}.
} 
\label{fig:Complementarity-plot2}
\end{figure}

As a second example, Fig.~\ref{fig:Complementarity-plot2} shows the constraints from $0\nu\beta\beta$ decay as well as from several collider processes, for 
a simplified scenario involving an additional scalar, fermion, and leptoquark field, for two different benchmark points of the model masses parameters~\cite{Graesser:2022nkv}. At low-energies, the leading 
contribution to $0\nu\beta\beta$ decay in this simplified model arises from so-called dimension-9 ``vector operators'', whose transition amplitude is 
chirally suppressed compared to the amplitude induced by other dimension-9 LNV operators. As a result and 
unlike the situation for the previous simplified model, here 
the 
LHC experiments have significantly better sensitivity compared to the $0\nu \beta \beta$ experiments, provided 
the LNV mediators can be produced at the LHC. The complementarity 
between these two experiments is additionally sensitive to the underlying mass spectra, as is evident from a cursory comparison of the two panels shown in Fig.~\ref{fig:Complementarity-plot2}. 
It should be noted that in making projections for exclusion and discovery at the HL-LHC, 
Ref.~\cite{Graesser:2022nkv}
makes the 
strong assumption
that the selection cuts and efficiencies remain unchanged from those in existing searches. The reaches presented there might be improved with an optimization of cuts and better understanding of 
backgrounds in the HL-LHC environment.  Moreover, 
signals of this model at the LHC and $0\nu\beta\beta$ experiments are entirely uncorrelated with the observed neutrinos masses, as these new sources of LNV give negligible contributions to the latter, arising as 
they do at three-loop order in this scenario.
\begin{figure}[!t] 
\centering
\includegraphics[width=0.55\linewidth]{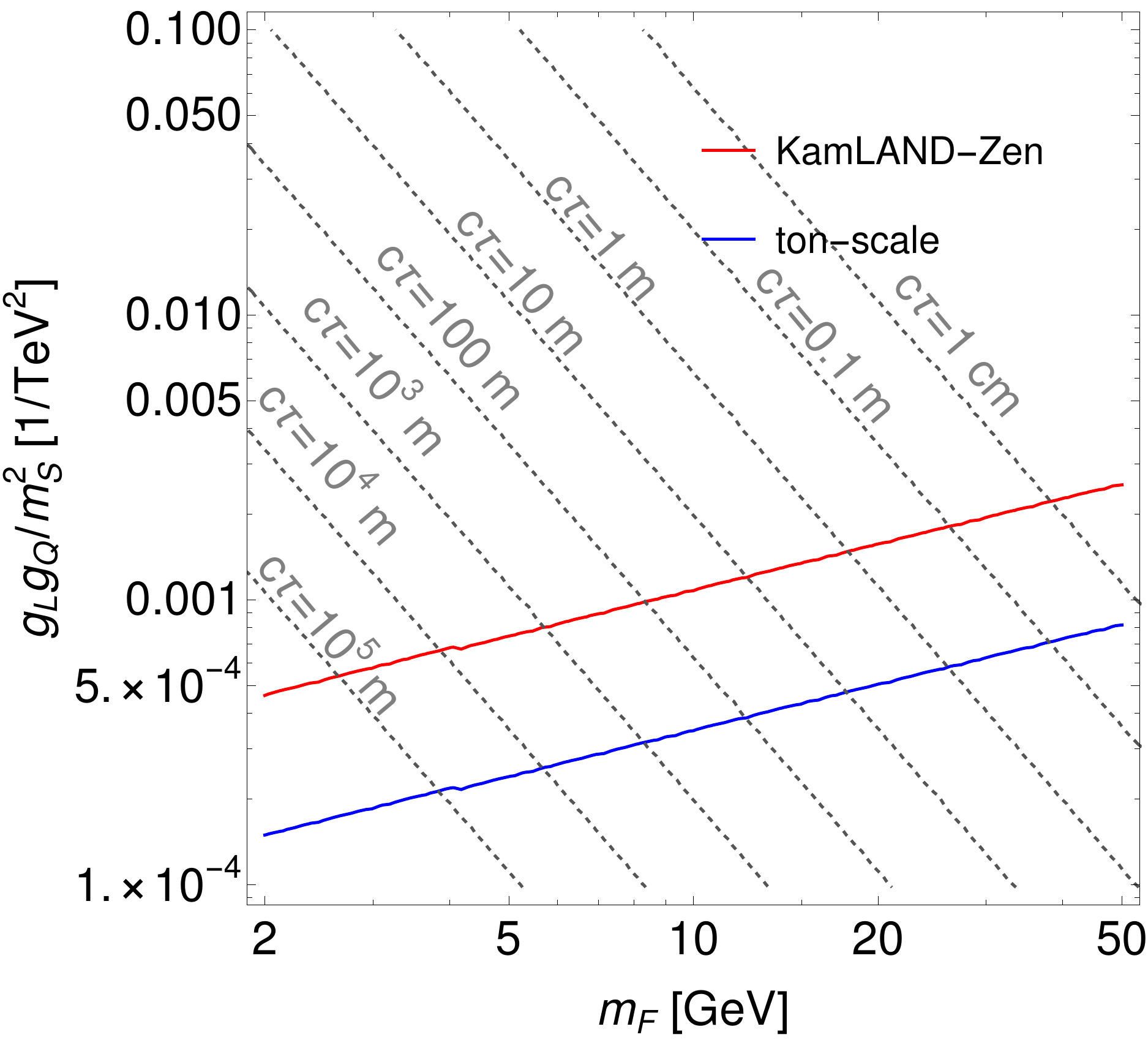} 
\caption{$0 \nu \beta \beta$  lifetime exclusion (red) from Kamland-Zen 
and projected limit (blue) from tonne-scale experiments, in a simplified model having a 
$SU(2)_L$ doublet scalar and a neutral Majorana fermion that is the LLP. Also overlaid in grey are 
the proper decay lengths of the LLP.
Figure is taken from Ref.~\cite{Li:2021fvw}. 
} 
\label{fig:Mathusla1}
\end{figure}
\begin{figure}[!t] 
\centering
\includegraphics[width=0.45\linewidth]{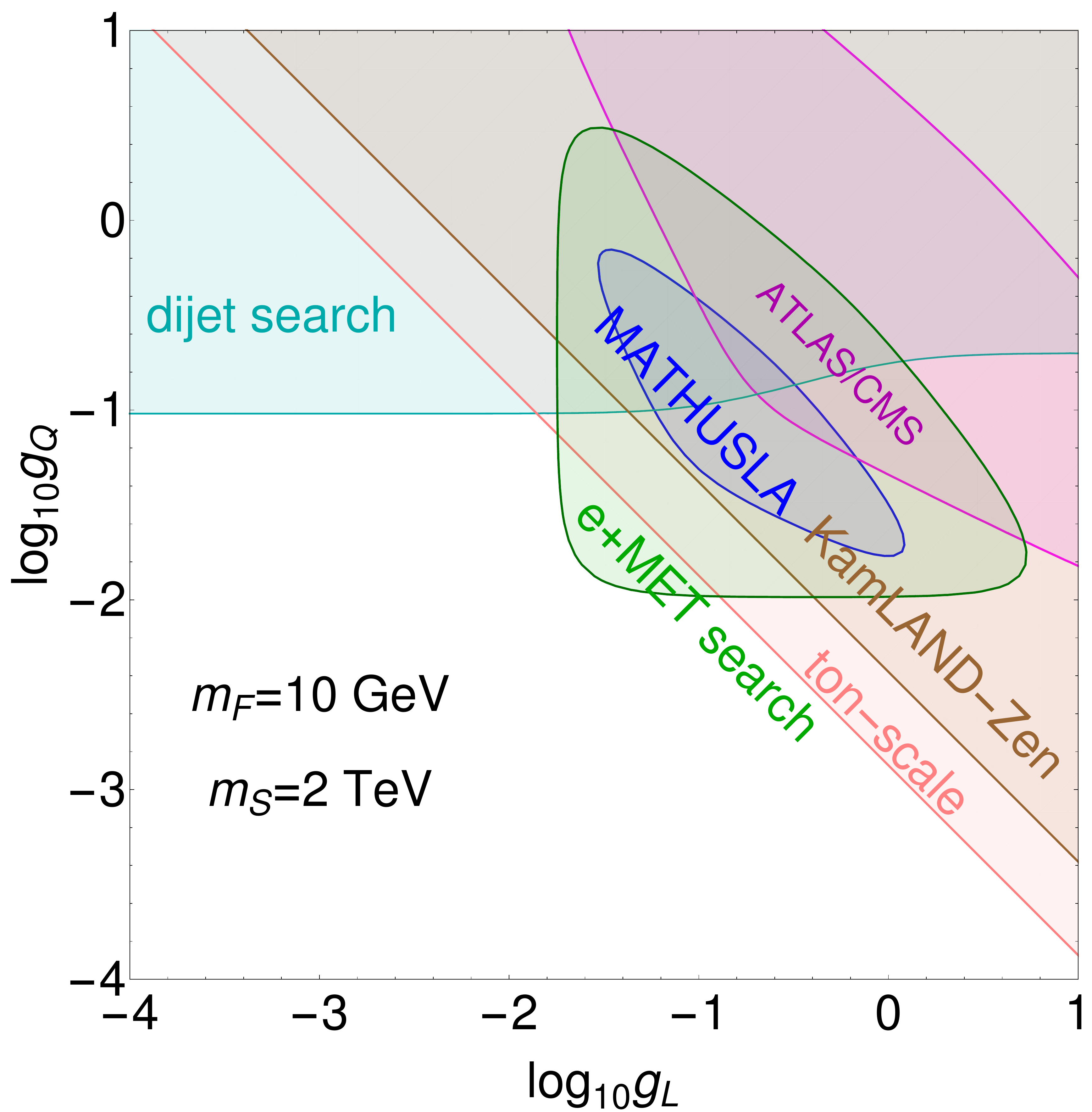}
\includegraphics[width=0.45\linewidth]{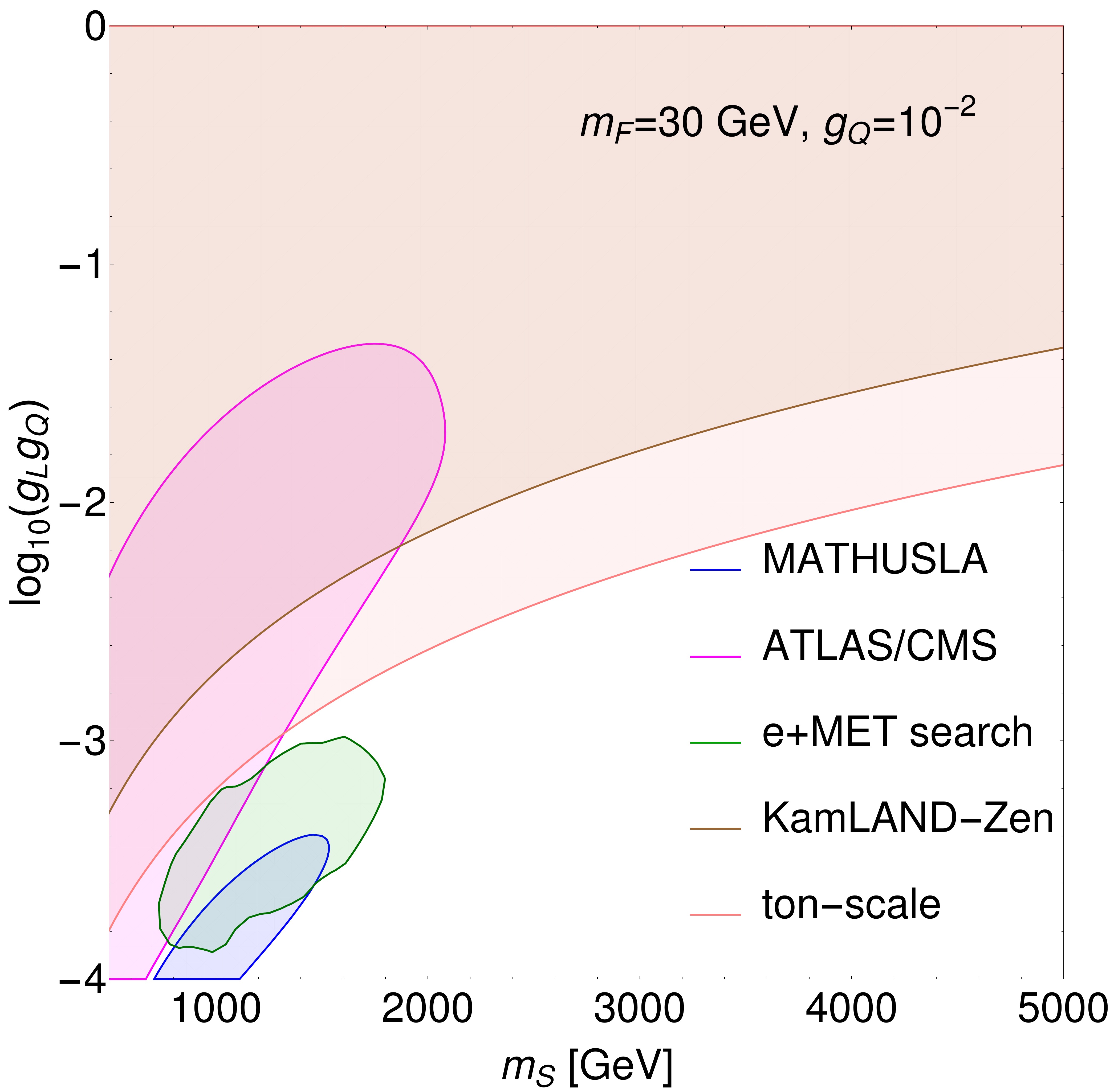}
\caption{The combined sensitivities in $0 \nu \beta \beta$ decay and the LHC in the
same simplified model used for Fig.~\ref{fig:Mathusla1}.
The shaded regions denote the current exclusion limit by the
KamLAND-Zen experiment (brown), and future tonne-scale 
$0 \nu \beta \beta$ decay experiments (pink). Also shown are  
LHC exclusion limits from dijet (cyan) and $e+$ missing energy (green) searches. 
Future LLP searches at the HL-LHC with $3$ ab$^{-1}$ of integrated luminosity are also shown for
ATLAS/CMS (magenta), and MATHUSLA (blue). The two panels correspond to two different benchmark points. 
Figures are taken from Ref.~\cite{Li:2021fvw}. 
} 
\label{fig:Mathusla2}
\end{figure}

Simplified models of TeV-scale LNV can also illustrate the interplay between not only $0\nu \beta \beta$ experiments and the ATLAS and CMS experiments, but also with proposed far detectors, such as FASER, 
SHiP, and MATHUSLA, among other experiments, that are dedicated to searching for long-lived particles (LLPs) that are produced at the primary interaction point and have macroscopic decay lengths~\cite{Alimena:2019zri,Feng:2022inv,Abdullahi:2022jlv}. LLPs can generically occur in models of TeV-scale LNV, in which there is a weakly coupled particle, typically a Majorana fermion, with a mass on the order of $\sim 10-50$ GeV. 
For instance, in simplified 
models with a light right-handed neutrino~\cite{Graesser:2007yj,Graesser:2007pc,Cottin:2018nms,Cottin:2018kmq}, the left-right symmetric model with a light right-handed neutrino~\cite{Nemevsek:2016enw,Cirigliano:2017djv,Nemevsek:2018bbt}, or in other simplified models with a light 
Majorana fermion~\cite{Li:2021fvw}, the $c \tau$ for such a light LLP can vary over a broad range, from $\sim 10^{-4}-10^5$ m. Production of {\em two} such LLPs, for instance in the decay of a single heavier particle or in the cascade decays of a pair of heavier particles, can lead to signals of overall LNV (such as same-signed dileptons) in the overall topology of the event. Such LNV 
would manifest in the aforementioned far detectors as a {\em pair} of decays (one from the decay of each LLP), which in aggregate violate overall lepton number. 

The current and projected sensitivities of the $0\nu \beta \beta$ experiments, compared to 
current and projected sensitivities of the LHC for a third simplified model of TeV-scale LNV are shown 
in Fig.~\ref{fig:Mathusla2}. Here, the simplified model contains a 
$SU(2)_L$ doublet scalar and a Majorana fermion. As shown in Fig.~\ref{fig:Mathusla1}, the 
Majorana fermion is a LLP over a broad range for its mass and its couplings to other particles. 
Combining 
searches for LLPs at the far detectors with searches for exotica at the main detectors clearly broadens the sensitivity of the LHC experiments to LNV, in ways that complements anticipated future limits from $0 \nu \beta \beta$ experiments~\cite{Li:2021fvw}.

\subsection{Neutrinoless Double-Beta Decay and Cosmology}
\label{sec:cosmo}
As the right panel of Fig.~\ref{fig:masses} illustrates, the $0\nu\beta\beta$ decay implications of astrophysical and cosmological probes of neutrino properties can be significant, in the absence of dynamics that depart from the standard neutrino thermal history. Conversely,  $0\nu\beta\beta$ decay and collider searches for LNV can yield important insights into the physics of the early universe.
In what follows, we illustrate each direction, focusing first on astrophysical bounds on $\Sigma$, and subsequently on the implications of discovering TeV-scale LNV for baryogenesis via leptogenesis.

Returning to the right panel of Fig.~\ref{fig:masses}, it is evident that future astrophysical probes of $\Sigma$ may rule out the inverted hierarchy band for $|m_{\beta\beta}|$ in the \lq\lq standard mechanism\rq\rq\, involving only three light Majorana neutrinos. The sensitivity of next generation tonne-scale $0\nu\beta\beta$ experiments will cover this band, implying significant discovery potential if nature has chosen this neutrino mass hierarchy. Should future astrophysical bounds on $\Sigma$ fall significantly below $10^{-1}$ eV, the remaining portion of the normal hierarchy band could be substantially reduced. In this context, it is worth observing that future cosmic microwave background (CMB) experiments could achieve a sensitivity of $\Sigma = 0.062\pm 0.016$~eV at the $1\sigma$ level~\cite{CORE:2016npo,Lattanzi:2017ubx}.
\begin{figure}[!t] 
\centering
\includegraphics[width=0.675\linewidth]{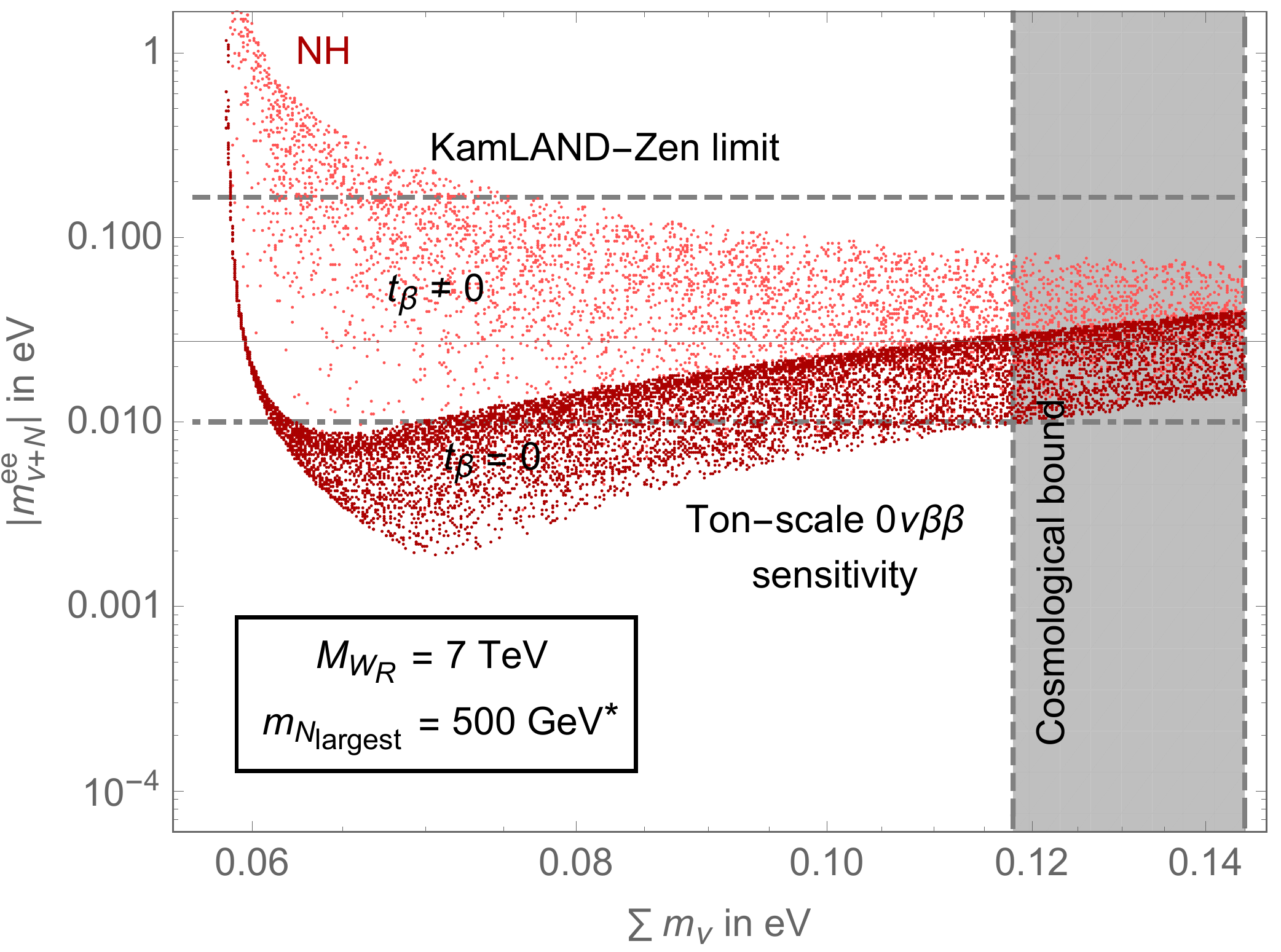}
\caption{$0\nu\beta\beta$ sensitivity in the minimal left-right symmetric model~\cite{Li:2020flq}. Vertical axis gives the effective mass as a function of the sum of three light neutrino masses. The grey region is excluded by current astrophysical constraints on the sum of light neutrino masses. Horizontal black dashed and dot-dashed lines give present $0\nu\beta\beta$ exclusion and future tonne-scale sensitivities, respectively. The light (dark) red points show the relation between the effective $0\nu\beta\beta$ mass $|m^{ee}_{\nu+N}|$ and $\Sigma$ in the normal hierarchy in the presence (absence) of mixing between the $W_L$ and $W_R$ bosons. Mixing is governed by the parameter $t_\beta$.} 
\label{fig:mLRSM}
\end{figure}

The presence of additional sources of LNV could alter the $0\nu\beta\beta$ implications of astrophysical constraints on $\Sigma$. To illustrate, we consider the presence of TeV-scale LNV in the context of the minimal left-right symmetric model. In this case, the situation for the inverted hierarchy remains largely unaltered from the standard mechanism. However, the $0\nu\beta\beta$ sensitivity to the normal hierarchy changes dramatically, particularly in the presence of mixing between left- and right-handed $W$ bosons, as shown in Ref.~\cite{Li:2020flq}. Figure~\ref{fig:mLRSM} shows the relation between $\Sigma$ and $|m_{\beta\beta}|$ (denoted $m^{ee}_{\nu+N})$ for the normal hierarchy. The light (dark) red points correspond to the presence (absence) of left-right mixing. Evidently, even in the presence of more severe constraints on $\Sigma$ from astrophysical probes, there exists considerable open parameter space that would accommodate a $0\nu\beta\beta$ signal. Importantly, a $0\nu\beta\beta$ observation in the context of such constraints would clearly point to new sources of LNV at or below the TeV scale --- a possibility that could be further elucidated with collider LNV searches.

The discovery of LNV at or below the TeV scale, while of great interest in its own right, could have profound implications for our understanding of the baryon asymmetry. In the type I seesaw mechanism that generates the dimension-5 Weinberg neutrino-mass operator, early universe CP-violating (CPV) decays of the heavy, right-handed Majorana neutrinos could provide the seeds for the observed baryon asymmetry via the well-known thermal leptogenesis scenario. While there exists no general, model-independent connection between this source of CPV and the CPV phase in the PMNS matrix, observation of the latter would perhaps increase the plausibility of this leptogenesis mechanism. Similarly, the observation of $0\nu\beta\beta$ decay would, under conventional type I seesaw mechanism, point to the existence of the Majorana neutrinos in the early universe, another key ingredient in the standard leptogenesis paradigm.
\begin{figure}[!t] 
\centering
\includegraphics[width=1\linewidth]{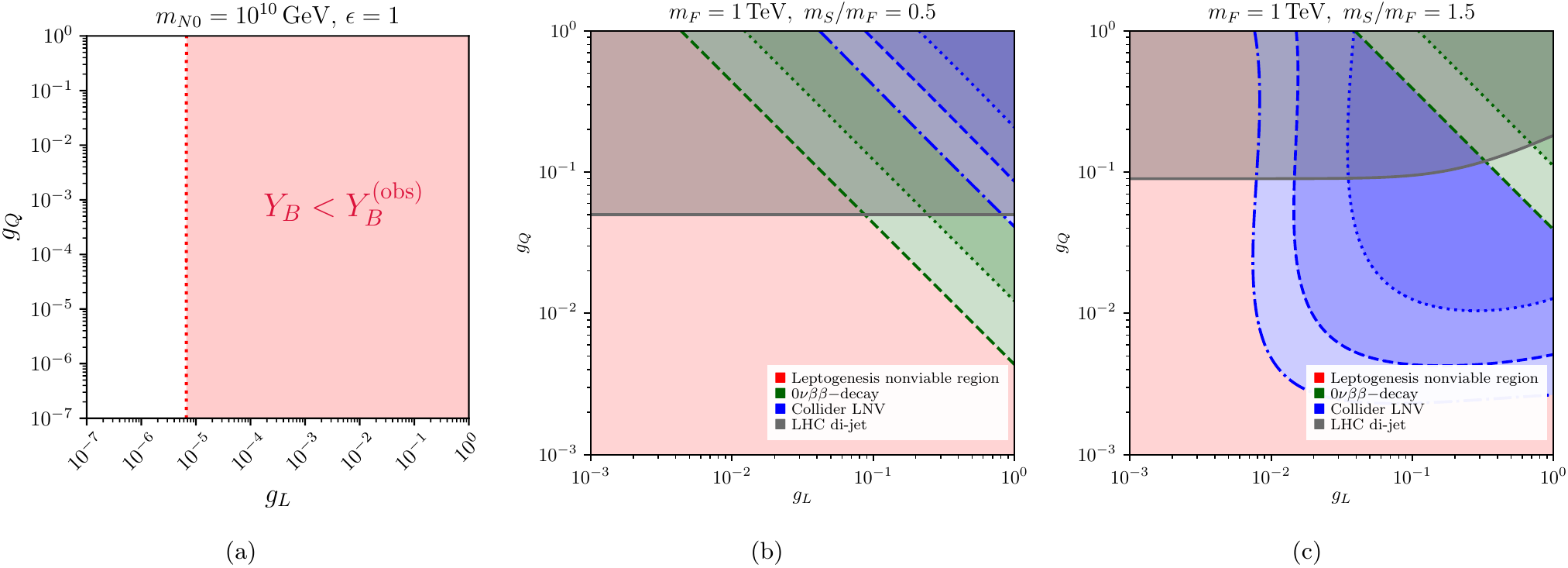}
\caption{Interplay of leptogenesis with $0\nu\beta\beta$ decay and collider sensitivity to TeV-scale LNV~\cite{Harz:2021psp}. In each panel, the pink region corresponds to a baryon asymmetry smaller than the observed value. In panel (a) the white region accommodates the observed asymmetry, assuming a maximal CPV asymmetry $\epsilon$ for heavy right-handed neutrinos. In (b) and (c), the grey shaded region is excluded by current LHC di-jet resonance searches. Present and tonne-scale $0\nu\beta\beta$ sensitivities are indicated by dark and light green regions, respectively. Present and prospective searches for a same-sign di-electron plus dijet search sensitivities are shown in blue for the LHC at 14 TeV with integrated luminosities of 100 fb$^{-1}$ (dark blue, dotted line) and 3 ab$^{-1}$ (medium blue, dashed line), and the FCC-hh at 100 TeV with 30 ab$^{-1}$ (light blue, dash-dotted line) . 
} 
\label{fig:lepto}
\end{figure}

The presence of additional sources of LNV at lower energy ({\it e.g.} TeV) scales, however,  could be fatal to the viability of leptogenesis. As observed based on general considerations and EFT methods in Refs.~\cite{Deppisch:2015yqa, Deppisch:2017ecm}, these additional LNV interactions could imply rapid $\Delta L=2$ and/or $\Delta L=1$ processes that would wash out a lepton  asymmetry (or even baryon asymmetry, if sphaleron processes are taken into account) generated at higher scales/early times by the CPV right-handed neutrino decays. Specifically, it was shown that, assuming the observation of $0\nu\beta\beta$ decay with a half-life around $10^{27}$ yr, 
a strong washout is triggered by 7-dimensional $\Delta L = 2$ SMEFT operators
arising at scales between $10^3$ and $10^5$ GeV, whereas  dimension-9 and -11
operators usually washout the asymmetry if $\Lambda$ is in between 100 GeV and 1 TeV.
Results of a simplified-model illustration~\cite{Harz:2021psp} appear in Fig.~\ref{fig:lepto}. The model extends the Standard Model with a new scalar doublet $S$ and singlet Majorana fermion $F$ whose interactions with the SM quarks and leptons are governed by couplings $g_Q$ and $g_L$, respectively. (The same model is used in obtaining Figs.~\ref{fig:Mathusla1} and \ref{fig:Mathusla2}). This study also assumes the presence of heavy right-handed Majorana neutrinos $N$ of the type I seesaw mechanism, whose out-of-equilibrium CPV decays provide a source for standard thermal leptogenesis.

In assessing the washout effects of the additional TeV-scale LNV interactions, the coupling $g_L$ is decisive. As indicated by the boundary between the pink and white regions in panel (a), the viability of standard thermal leptogenesis would require $g_L\lesssim 10^{-5}$. On the other hand, the sensitivity of tonne-scale $0\nu\beta\beta$ searches, as well as that of prospective future collider probes (assuming prompt decays), falls well within the pink region that cannot accommodate the observed baryon asymmetry, as indicated in panels (b) and (c). Note also that the relative reaches of  $0\nu\beta\beta$ decay, the LHC, and a future 100 TeV $pp$ collider depend on the new-particle spectrum. Thus, the presence or absence of a signal in $0\nu\beta\beta$ experiments and $pp$ collisions provides a diagnostic of the model and its cosmological implications.

\subsection{Low-Energy EFTs for Lepton Number Violation}
\label{sec:EFTs-for-LNV}
\begin{figure}[!t]
    \centering
    \includegraphics[width=\textwidth]{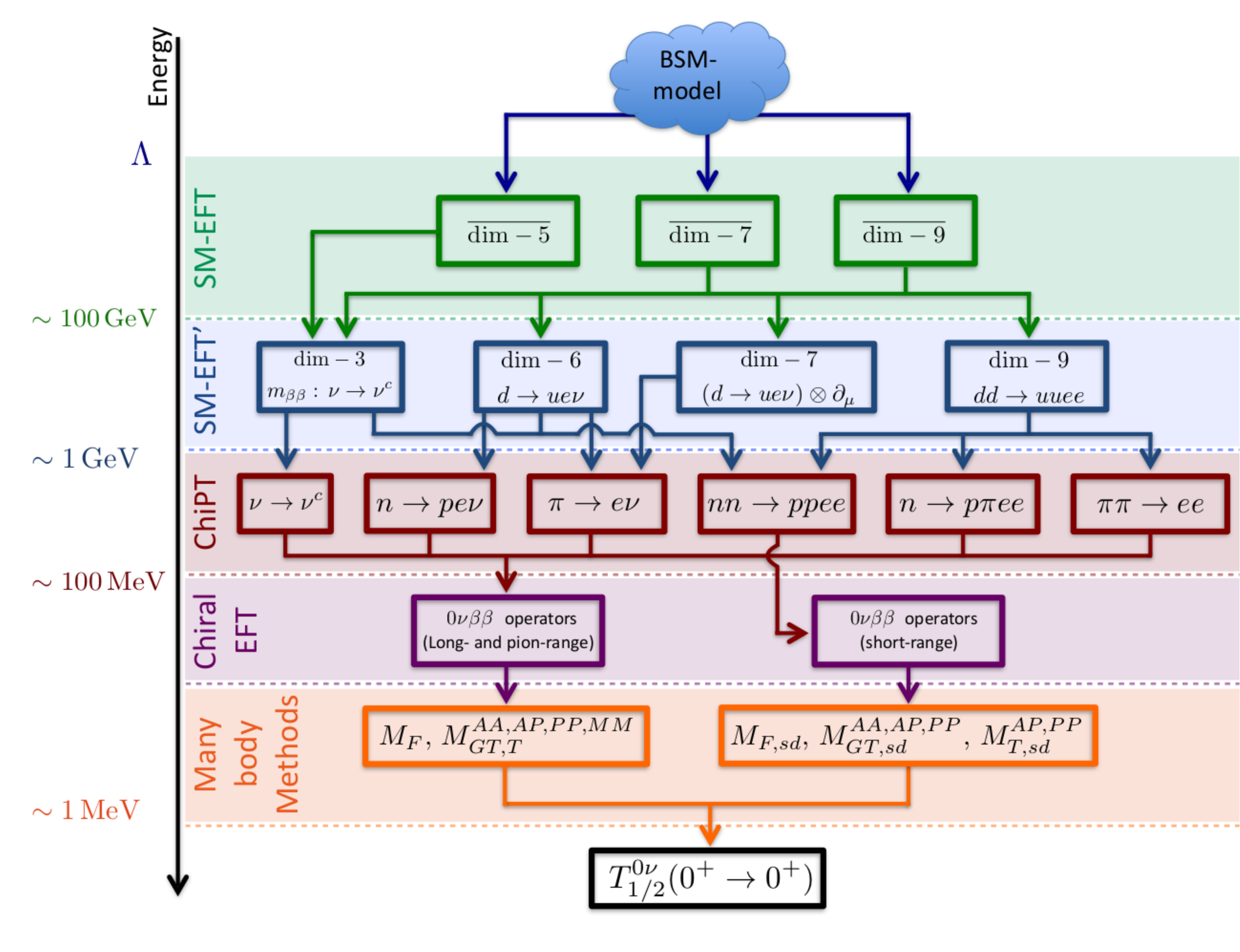}
    \caption{The tower of EFTs for $0\nu\beta\beta$ decay. At the electroweak scale, LNV operators are described by operators of odd dimension in the SMEFT. Heavy SM degrees of freedom can be integrated out by matching SMEFT onto LEFT (denoted as SMEFT$^\prime$ in the figure). Quark-level operators are then matched onto hadronic operators. The construction of hadronic operators is performed in $\chi$PT and chiral EFT, while the determination of the low-energy couplings requires non-perturbative techniques, such as lattice QCD. The $0\nu\beta\beta$ transition operators constructed in chiral EFT are then evaluated with nuclear many-body methods.
    Figure is adapted from Ref.~\cite{Cirigliano:2018yza}.}
    \label{fig:landscape}
\end{figure}
If the scale of new physics $\Lambda$ is larger than the electroweak scale $v$,
contributions to $0\nu\beta\beta$ decay can be organized in terms of effective LNV operators that 
only contain SM fields and
are invariant under the gauge symmetries of the Standard Model. The tower of effective operators is organized according to their scaling in powers of $v/\Lambda$. The EFT approach is schematically represented in Fig.~\ref{fig:landscape}, and consists of a series of matching steps in which degrees of freedom of 
virtualities much larger than the typical nuclear scales are progressively integrated out. 
This approach has the advantage that the
various steps connecting physics at and above the electroweak scale to $0\nu\beta\beta$ experiments --- including renormalization-group evolution to low-energy, the matching between quark- and hadron-level theories (``hadronization"), and the calculation of the NMEs of $0\nu\beta\beta$ transition operators, see Fig. \ref{fig:landscape} --- can be done once and for all, independently of the BSM model. Furthermore, each matching step has, in principle, a controllable theoretical uncertainty, determined by missing terms in the expansion in $\Lambda_{\rm low}/\Lambda_{\rm high}$ and in perturbative SM couplings. 

Starting at the electroweak scale, 
in the absence of additional light degrees of freedom besides those in the Standard Model, gauge-invariant and 
baryon-number conserving LNV operators
only occur for operators having an odd dimension~\cite{Kobach:2016ami}, and start at dimension~5~\cite{Weinberg:1979sa}, where a single operator can be written and, after electroweak symmetry breaking, gives rise to a Majorana mass
of order $m_\nu \sim v^2/\Lambda$. 
The smallness of neutrino masses can thus be explained by taking $\Lambda$ to be 
very high, $\Lambda \sim 10^{14}$ GeV. If, on the other hand, $\Lambda = \mathcal O({\rm TeV})$, the dimension-5 operator needs to be suppressed by small couplings, which can be achieved by various mechanisms~\cite{Mohapatra:1979ia,Zee:1980ai,Zee:1985id}. In this case, dimension-7 and dimension-9 operators also become important.

In the last few years, complete sets of dimension-7~\cite{Babu:2001ex,Bell:2006wi,deGouvea:2007qla,Lehman:2014jma,Liao:2016hru} and dimension-9~\cite{Prezeau:2003xn,deGouvea:2007qla,Graesser:2016bpz,Li:2020xlh} operators have been derived. These operators are sufficient to capture the great majority of BSM models, including the left-right symmetric model~\cite{Mohapatra:1979ia}, leptoquark ~\cite{Hirsch:1996ye}, and supersymmetric models~\cite{Hirsch:1995zi}. New light degrees of freedom, such as sterile neutrinos, can also straightforwardly be included in the Standard Model Effective Field Theory (SMEFT), and  
the full set of LNV operators with sterile neutrinos is known up to dimension~7~\cite{Liao:2016qyd}.

Heavy SM degrees of freedom can then be integrated out, and the SMEFT operators, with or without sterile neutrinos, can be matched to a $SU(3)_c \times U_{\rm em}(1)$ EFT (low-energy EFT or LEFT).  
The matching has been worked out for dimension-7 interactions and most dimension-9 operators, at tree level and including QCD evolution at leading logarithms~\cite{Cirigliano:2017djv,Cirigliano:2018yza,Graf:2018ozy,Deppisch:2020ztt}. The next step down in energy consists in matching the LEFT onto theories with hadronic degrees of freedom, which can then be used to derive $0\nu\beta\beta$ transition operators and calculate NMEs. 
For the dimension-5 Weinberg operator, this step has a long history 
\cite{Furry:1939qr,Haxton:1984ggj,Doi:1985dx,Simkovic:1999re}.
For higher-dimensional operators, the hadronization was carried out in specific models~\cite{Doi:1985dx,Hirsch:1996ye,Faessler:2007nz}. A first EFT-based approach was developed in Refs.~\cite{Pas:1999fc,Pas:2000vn} and later reviewed in Ref.~\cite{Deppisch:2012nb}, which however relied on a factorization assumption 
to parametrize nucleon-level operators. 

The hadronic representation of 
LNV LEFT operators 
can be constructed more rigorously by using low-energy EFTs of QCD,
which encode the symmetries of QCD, in particular chiral symmetry, and organize hadronic interactions in a systematic expansion in $\epsilon_\chi = Q/\Lambda_\chi$, with
$\Lambda_\chi \sim 1$ GeV and $Q$ denoting low-energy scales of order of the pion mass. 
In the mesonic and single-nucleon sector, interactions can be constructed in
chiral perturbation theory ($\chi$PT)
\cite{Weinberg:1978kz,Gasser:1983yg,Jenkins:1990jv}. 
The extension of $\chi$PT to the multi-nucleon sector is often called chiral EFT~\cite{Weinberg:1991um}. The power counting in
chiral EFT is
significantly more complicated due to the non-perturbative nature of nuclear forces, which often requires to modify the naive scaling of contact interactions in order to obtain results that are explicitly independent of regulators introduced to solve the nuclear few- and many-body problem~\cite{Kaplan:1996xu,Nogga:2005hy}.

The application of $\chi$PT and chiral EFT ideas to $0\nu\beta\beta$ decay was pioneered in
Ref.~\cite{Prezeau:2003xn}, which performed the first systematic study of the hadronization of dimension-9 operators and of the construction of the transition operator in $\chi$PT, and
pointed out the dominance of pionic contributions for several dimension-9 operators.
In the case of specific models (R-parity-violating supersymmetry) the importance of pion interactions was also noticed in Ref.~\cite{Faessler:1996ph}.
Continuing the path of Ref.~\cite{Prezeau:2003xn}, Ref.~\cite{Graesser:2016bpz} consistently applied chiral EFT with Weinberg's power counting to
assess the order at which a long-distance pion enhancement first appears in the chiral expansion of dimension-9 operators, reproducing and extending to next-to-next lowest
order the power counting results of Ref.~\cite{Prezeau:2003xn}.
A derivation 
of the chiral EFT realization of dimension-5, -7 and -9 operators was carried out in Refs.~\cite{Cirigliano:2017djv,Cirigliano:2017tvr,Cirigliano:2018yza},
with a study of the internal consistency of the transition operators induced by long-range neutrino and pion exchanges performed in Refs.~\cite{Cirigliano:2018hja,Cirigliano:2018yza,Cirigliano:2019vdj}.
These works were extended to SMEFT operators involving light sterile neutrinos in Ref. \cite{Dekens:2020ttz}.
\begin{figure}[!t]
    \centering
    \includegraphics[width=0.8\textwidth]{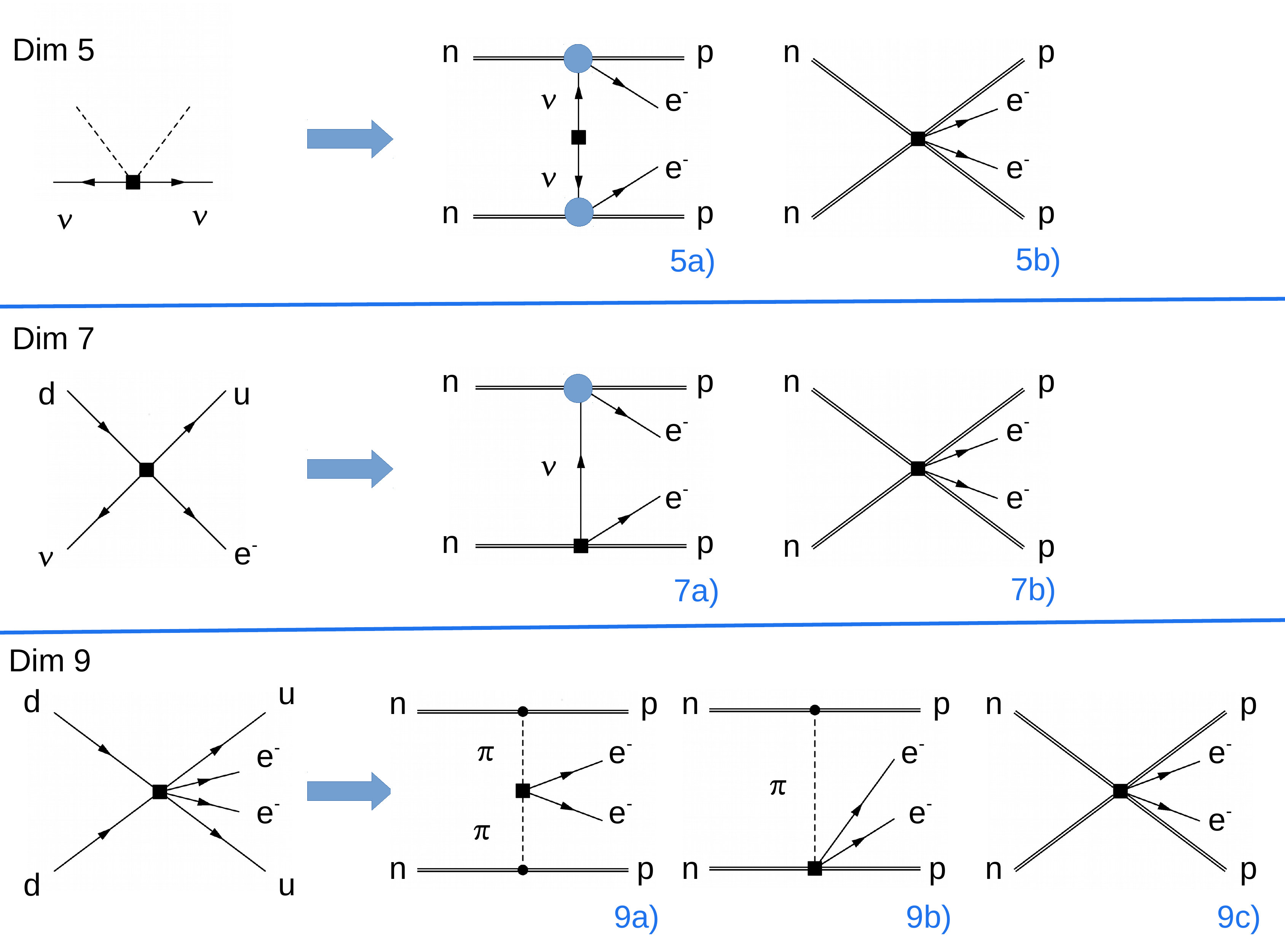}
    \caption{Matching between effective operators in the LEFT and chiral EFT. The left-hand side denotes representative dimension-5, -7 and -9 LNV operators in the LEFT, after integrating out heavy degrees of freedom. Plain lines denote quarks and leptons, while dashed lines Higgs fields. When matching onto LEFT, the Higgs fields are, in most cases, replaced by the Higgs vacuum expectation value. The diagrams on the right-hand-side denote leading contributions to the $0\nu\beta\beta$ operators.
    Double, dashed and plain lines denote nucleons, pions and leptons, respectively. The blue bubbles denote nucleon form factors, which subsume pion-range contributions to the pseudoscalar form factor and to the induced pseudoscalar component of the axial form factor.
    LNV interactions are denoted by a black square.
    }
    \label{fig:matching}
\end{figure}
The matching between quark-level and hadron-level operators is schematically illustrated in Fig.~\ref{fig:matching}.

Dimension-5 and -7 operators typically induce LNV interactions involving light neutrinos, with LNV (denoted by a black square in Fig.~\ref{fig:matching}) arising from the neutrino Majorana mass or from LNV interactions of leptons with quark bilinears. In chiral EFT, these interactions induce LNV couplings of leptons to pions and to nucleon bilinears and are proportional to well-known low-energy couplings, such as the pion decay constant, the quark condensate or the nucleon axial, scalar and tensor charges. The exchange of neutrinos between nucleons then gives rise to a long-range component in the $0\nu\beta\beta$ transition operator,
shown by the first diagrams on the right-hand-side of Fig.~\ref{fig:matching}.
Refs.~\cite{Cirigliano:2018hja,Cirigliano:2018yza,Cirigliano:2019vdj}
pointed out that the matrix elements of the long-range transition operators are often regulator dependent, and require 
LNV contact interactions, without explicit neutrinos, to yield well defined results.
Dimension-9 operators (last row in Fig.~\ref{fig:matching}) give rise
to local LNV interactions of
two charged leptons with four quarks, which in the EFT generate couplings of pions and nucleons to two same-sign leptons.

In general then,
for a given LNV SMEFT operator, 
the $0\nu\beta\beta$ transition operator 
will have both a long-range (Coulomb-like or pion-range) and a short-range component. 
 In order to keep a systematic connection with neutrino and BSM models, the coupling constants entering these components need to be determined from first principles. The
 state of the art can be summarized as follows:
\begin{itemize}[leftmargin=*]
    \item[$\circ$] \emph{Coulomb-range components (diagrams $5a$ and $7a$ in Fig.~\ref{fig:matching}).} The hadronic input is determined by the single-nucleon vector, axial, scalar, pseudoscalar and
    tensor isovector form factors, which can be extracted from data or have been calculated with good accuracy in lattice QCD~\cite{Aoki:2021kgd};
    \item[$\circ$] \emph{Pion-range components (diagrams $9a$ and $9b$).} The $\pi\pi$ couplings induced by dimension-9 operators have been obtained via dedicated lattice-QCD calculations, with $\sim \mathcal O(10\%)$
    uncertainties~\cite{Nicholson:2018mwc}.
    Progress has also been made on $\pi\pi$
    couplings induced by light-neutrino exchange~\cite{Feng:2018pdq,Tuo:2019bue,Detmold:2020jqv}, which contribute at next-to-next-to-leading order (N$^2$LO) in chiral EFT. The $\pi N$ couplings entering diagram $9b$ need to be determined;
    \item[$\circ$] \emph{Short-range components (diagrams $5b$, $7b$ and $9c$).} 
    The $nn \rightarrow p p e^- e^-$ contact interaction induced by light-Majorana exchange, denoted by $g_\nu^{\rm NN}$
    in Ref.~\cite{Cirigliano:2018hja,Cirigliano:2019vdj},
    has so far been determined via large-$N_c$ considerations 
    \cite{Richardson:2021xiu} and dispersive methods~\cite{Cirigliano:2020dmx,Cirigliano:2021qko} inspired by the Cottingham formula for the electromagnetic contributions to the nucleon mass~\cite{Cottingham:1963zz}. All other leading-order $nn \rightarrow pp$ couplings are  unknown beyond naive factorization estimates.
\end{itemize}
The strength of $nn \rightarrow pp$ couplings induced by dimension-5, -7 and -9 operators is an important source of uncertainty in $0\nu\beta\beta$ NMEs~\cite{Wirth:2021pij,Jokiniemi:2021qqv}
 and thus in the extraction of $|m_{\beta\beta}|$ and/or of LNV parameters in BSM theories. For this reason, in the last few years the lattice-QCD community has undertaken an effort for the calculation of these couplings, which will be described in the next section.
 
Finally, Fig.~\ref{fig:matching}
captures the leading-order contributions in chiral EFT. One of the advantages of chiral EFT is that it allows for the construction and estimate of subleading operators. The $0\nu\beta\beta$ transition operators receive corrections at N$^2$LO, which should affect NMEs at the $\sim$ 10\% level.
At N$^2$LO, one encounters loop corrections to the two-body transition operators~\cite{Cirigliano:2017tvr}, momentum-dependent 
short-range interactions~\cite{Cirigliano:2019vdj},
closure corrections, 
and corrections arising from two-body weak currents~\cite{Menendez:2011qq}. 
Up to now, higher-order corrections have been considered only for the standard light-neutrino exchange mechanism, and only in the  Weinberg's power counting of chiral EFT. 
In the broad effort of establishing a rigorous strategy for quantifying the uncertainties of
theoretical predictions of $0\nu\beta\beta$
half-lives, one of the goals of the nuclear EFT community in the coming years is to systematize the construction of subleading corrections to $0\nu\beta\beta$ transition operators, and validate the convergence of the chiral expansion for $0\nu\beta\beta$ candidates.

\section{Lattice-QCD Calculations for Neutrinoless Double-Beta Decay}
\noindent
Lattice-QCD methods enable calculations of few-nucleon quantities of relevance for $0\nu\beta\beta$ processes. Given the significant computational cost of lattice-QCD calculations, direct calculations of $0\nu\beta\beta$ processes in experimentally relevant nuclei from QCD are not computationally feasible. Thus, matching few-nucleon matrix elements calculated from lattice QCD to EFTs will be required to connect to experiment. Furthermore, due to the restriction of lattice-QCD calculations to a finite Euclidean spacetime, extracting physical observables may also require the use of EFTs. 

Currently, the most outstanding and computationally feasible results needed from lattice QCD are related to $nn \to pp$ processes. The necessary lattice-QCD calculations for such processes can be broken down into roughly three stages: calculation of two-nucleon spectra and elastic scattering amplitudes, including developing the necessary operators for two-nucleon matrix elements, calculation of the finite-volume two-nucleon matrix elements of interest, and the extraction of physical infinite-volume, Minkowski-space transition amplitudes. These aspects of lattice-QCD calculations related to $0\nu\beta\beta$ decay are discussed in the following, along with avenues for future investigations. 

\subsection{Two-Nucleon Calculations in Lattice QCD
\label{sec:NN-LQCD}}
The current goal from the lattice-QCD perspective is to calculate the $nn\to pp$ transitions for a variety of operators. However, these processes can not be attempted until the more fundamental $NN \to NN$ scattering processes have been understood within the lattice-QCD framework. Lattice-QCD calculations of two-nucleon scattering are necessary for ensuring that: a) operators that couple well to the physical two-nucleon systems are identified for the use in $nn\to pp$ processes; b) systematic uncertainties associated with the lattice-QCD calculations in the $NN$ sector are understood and sufficiently controlled; and c) infinite-volume transition amplitudes can be extracted from finite-volume matrix elements, a process that requires access to $NN$ finite-volume energies and energy-dependence of the $NN$ matrix element, as discussed in Sec.~\ref{sec:extracting}.

Recent work within the lattice-QCD community has highlighted the importance of fully understanding the $NN$ spectra. There remains an open question surrounding the existence of bound states in two-nucleon systems at unphysically large quark masses (see, e.g., Ref.~\cite{Drischler:2019xuo} for a review of this issue). This debate had largely focused on the use of different methods for extracting the scattering phase shifts, whether through the use of a model-independent framework known as L\"uscher formalism~\cite{Luscher:1986pf,Luscher:1990ux}, or the HALQCD potential method~\cite{Ishii:2012,Aoki:2012tk,Aoki:2012bb}, with the former predicting bound states in both s-wave $NN$ channels~\cite{NPLQCD:2013bqy,Berkowitz:2015eaa,Wagman:2017tmp,Orginos:2015aya,NPLQCD:2020lxg,Yamazaki:2012hi,Yamazaki:2015asa} and the latter no bound state~\cite{Inoue:2013nfe,Murano:2013xxa,Ishii:2012,Aoki:2012tk,Aoki:2012bb}. However, this debate has gradually shifted to the operators used in the two approaches, as calculations in a given method seemed to show discrepancies in the extracted ground-state energies for different operators~\cite{Iritani:2018vfn}. More recently, the use of variational techniques for diagonalizing a large operator basis has found energies that are several standard deviations higher than previous estimates, implying that there is either no deep bound state in the spectrum at these large quark masses or that the interpolating-operator set is still not sufficiently complete to overlap significantly onto such states if present~\cite{Green:2021qol,Francis:2018qch,Horz:2020zvv,Amarasinghe:2021lqa}. 

These studies have been 
predominantly
carried out at very large quark masses, where extrapolation to the physical point cannot be controlled. Moving towards the physical point makes the resolution of the nature of $NN$ spectrum even more challenging.
This is due to the infamous signal-to-noise issue, in which statistical fluctuations grow exponentially with both the lighter pion mass, larger atomic number, and larger Euclidean time~\cite{Parisi:1983ae,Lepage:1989hd,Beane:2009gs}. The growth with Euclidean time can be particularly insidious because if a sub-optimal two-nucleon operator set is used, the ground state is not reached until very large Euclidean times, long after the noise has overwhelmed the signal. This highly-correlated noise can obscure slow-time evolution, leading to underestimated systematic uncertainties. While improvements to the coupling to the low-lying states through a sufficient operator basis will be crucial to extracting reliable results, this issue might be further alleviated through a careful analysis of excited-state contributions at earlier Euclidean times, where the data is more precise. 

Most lattice-QCD calculations of two-nucleon scattering to date have been performed only at a single lattice spacing, as the statistical noise encountered in two-nucleon calculations was considered the more daunting obstacle, and furthermore, discretization effects in the extracted spectra for other systems have been found to be relatively modest (see, e.g., Ref.~\cite{Wilson:2019wfr}). Recently, however, the first test of the continuum limit of the binding energies of two-baryon systems at large quark masses have been performed, where it was found that discretization effects could shift significantly the binding energy of the H-dibaryon  away from the continuum value~\cite{Green:2021qol}. Further tests of discretization effects are thus warranted to determine whether this finding is relatively universal, i.e., it holds for other two-baryon systems and other lattice actions, if these large discretization effects persist at smaller values of the quark masses, or if these findings are caused by some additional correlated systematics.

Lattice-QCD studies of $NN$ systems must also be carried out at quark masses which are sufficiently close to the physical point so that reliable extrapolations may be performed. These extrapolations rely on the use of EFTs, which in turn rely upon lattice-QCD data to determine the relevant LECs. Thus, an interplay between lattice-QCD calculations of two-nucleon observables and EFTs will be necessary to determine at which quark masses one may be able to trust results for $0\nu\beta\beta$ quantities.

\subsection{Neutrinoless Double-Beta Decay Operators and Matrix-Element Techniques}
\label{sec:operators-and-matrix-element-techniques}
\begin{figure}[!t]
    \centering
    \includegraphics[scale=0.795]{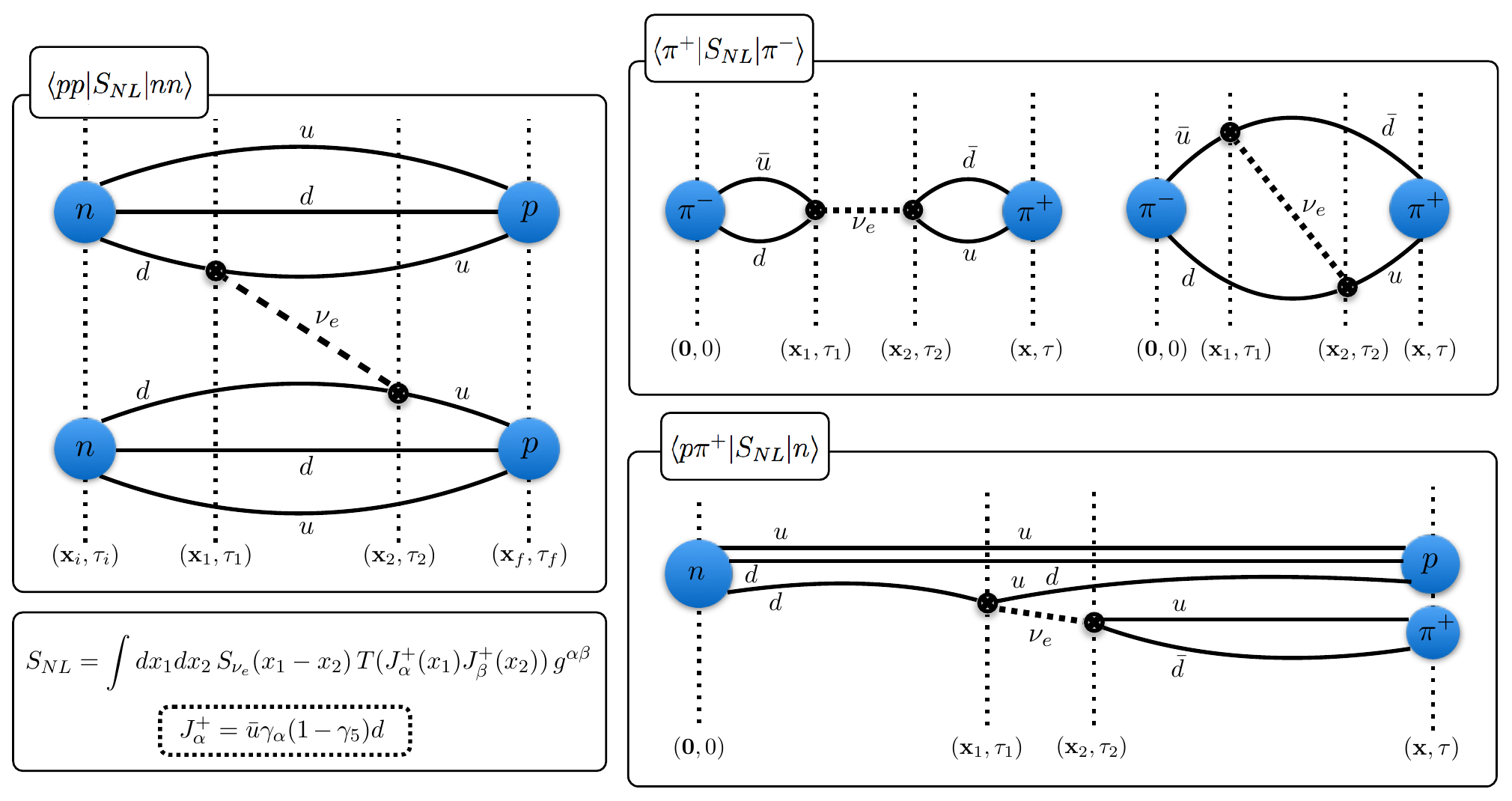}
    \caption{Diagrammatic representation of the four-point correlation functions that need to be evaluated to access $nn \to pp$, $\pi^- \to \pi^+$, and $n \to p \pi^+$ matrix elements in the light Majorana-neutrino exchange scenario. $S_{NL}$ denotes a non-local extended current, containing the neutrino propagator $S_{\nu_e}$, whose matrix elements between given hadronic states are obtained from the correlation functions shown. Figure is taken from Ref.~\cite{Davoudi19}.}
    \label{fig:LQCD-NL-ME}
\end{figure}
The diagrams that need to be constructed and calculated with the lattice-QCD method are illustrated in Fig.~\ref{fig:LQCD-NL-ME} for the $0\nu\beta\beta$ processes mediated by light Majorana neutrino exchange, where the light Majorana-neutrino propagator is analytically known.
These diagrams are similar to the QED corrections to hadronic correlation functions, with the photon propagator replaced by a neutrino propagator.
Due to this similarity, treatments of the divergence of the zero mode of the photon propagators
and the potential large finite-volume effects in QED can be applied to the $nn \to pp$ matrix-element calculations as well.
Possible methods include
removing all spatial zero modes (QED$_L$)~\cite{Blum:2007cy,Hayakawa:2008an,Borsanyi:2014jba,Davoudi:2018qpl},
introducing a non-zero photon mass~\cite{Endres:2015gda},
the infinite-volume photon reconstruction method (IVR)~\cite{Feng:2018qpx}, among other methods.
With these different form of neutrino propagators, the analysis that is needed
to relate the correlation function calculated with lattice QCD and the corresponding
infinite-volume matrix elements must be adjusted accordingly, as described in Sec.~\ref{sec:extracting}.

However, there is an important difference between $0\nu\beta\beta$ and QED corrections to hadronic amplitudes.
For the $0\nu\beta\beta$ process,
the neutrino-quark vertices change the quark flavor from down to up and vice versa.
Therefore, two interactions of this type can not be inserted on the same quark line.
This feature reduces the possible diagrams in the $0\nu\beta\beta$ processes
compared with the QED corrections to hadronic correlation functions.
Taking advantage of this feature of the diagrams, there are
two approaches to calculate the relevant matrix elements at present:
\begin{itemize}[leftmargin=*]
    \item[$\circ$] Calculate quark propagators with sources at both the hadrons' creation
    and annihilation locations, then perform contractions,
    sum over the entire space-time volume for both vertices,
    and include the neutrino propagator exactly with its analytical form.
    The volume-square cost needed for the naive summation for both vertices
    can be avoided by fast Fourier transformation techniques~\cite{Tuo:2019bue,Detmold:2020jqv}.
    \item[$\circ$] Introduce stochastic neutrino fields, calculate the quark propagators
    with sources at the hadrons' creation locations, then calculate sequential propagators with the stochastic neutrino field.
    The matrix elements can be obtained by performing contractions at the hadrons'
    annihilation locations using the quark propagators obtained in the previous steps.
\end{itemize}

Both approaches have their advantages.
For the first approach,
the neutrino propagators can be included without any added stochastic noise.
Also, different forms of the neutrino propagators
can be accommodated simultaneously without needing additional quark propagators.
Furthermore, the quark propagators required for this approach can be useful for many different projects.
The pion matrix elements in the upper right plot in Fig.~\ref{fig:LQCD-NL-ME}
have been calculated this way successfully~\cite{Tuo:2019bue,Detmold:2020jqv}.
The calculations can already be performed at physical pion masses.
Continuum extrapolation and finite-volume effects are also studied.
Further increase in the precision and better control of systematic effects
can be done without much difficulties.
To apply this approach to the proton/neutron matrix-elements calculation,
the number of diagrams and the complexity of contractions will grow significantly.
This will be a significant challenge in terms of coding complexity
and computational cost, but progress in this computation is underway.

For the second approach, more quark propagators are needed, as different sets of sequential quark propagators should be used if one implements different forms of the neutrino propagators (e.g. different neutrino masses to control finite-volume effects).
Compared with the first approach, additional statistical errors will be introduced from the stochastic neutrino field.
On the other hand, the contraction complexity and cost is significantly reduced.
One can simply reuse the same contraction code for the corresponding hadron correlation functions without
weak-interaction vertices, only replacing quark propagators by combinations of the original quark propagators and sequential quark propagators.
\begin{figure}[!t]
    \centering
    \includegraphics[scale=0.6]{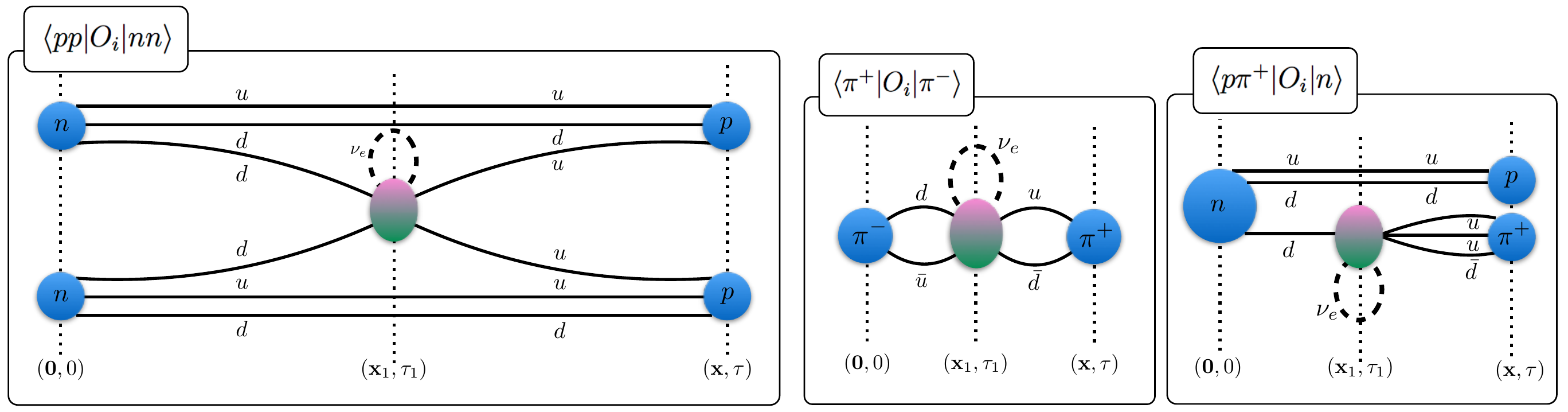}
    \caption{Diagrammatic representation of the three-point correlation functions that need to be evaluated to access $nn \to pp$, $\pi^- \to \pi^+$, and $n \to p \pi^+$ matrix elements in a high-scale scenario. $O_i$ denotes dimension-9 four-quark-two-lepton operators of the SMEFT described in Sec.~\ref{sec:EFTs-for-LNV}, whose matrix elements between given hadronic states are obtained from the correlation functions shown. Figure is taken from Ref.~\cite{Davoudi19}.}
    \label{fig:LQCD-SR-ME}
\end{figure}

Another important scenario is the $0\nu\beta\beta$ processes
through local dimension-9 operators shown in Fig.~\ref{fig:matching}.
The diagrams that need to be constructed and computed with lattice QCD are illustrated
in Fig.~\ref{fig:LQCD-SR-ME}.
These diagrams can be viewed as a special case of the diagrams shown in Fig.~\ref{fig:LQCD-NL-ME}
where the neutrino propagators reduces to a point.
This scenario involves calculation of matrix elements of a number of known local operators $O_i$~\cite{Prezeau:2003xn} in the two-nucleon sector. These calculationally are not too different from several standard processes such as single-beta decay, $pp$-fusion, and hadronic parity violation in the two-nucleon sector, for which initial computations at large quark masses have been performed in recent years~\cite{Savage:2016kon,Kurth:2015cvl}.
Finally, compared with the light Majorana-neutrino case,
the local case has an additional complexity---the dimension-9 operators in the lattice-QCD calculation need proper
renormalization to match with the continuum operators.
This is commonly obtained through non-perturbative renormalization. The matrix elements associated with the $\pi^- \to \pi^+$ process are evaluated for this scenario~\cite{Nicholson:2018mwc}. These enter the EFT construction of the $nn \to pp$ matrix elements at given orders in the chosen power-counting scheme~\cite{Prezeau:2003xn,Cirigliano:2018yza}, and are of importance in the EFT program for $0\nu\beta\beta$ decay.

These examples demonstrate how the matrix elements of relevance to $0\nu\beta\beta$ decay in the few-hadron sector are calculated at an operational level, and how they may be extended to evaluate other closely-related matrix elements, such as those involving dimension-7 operators, $\pi N$ LNV matrix elements (see Figs.~\ref{fig:LQCD-NL-ME} and \ref{fig:LQCD-SR-ME}), as well as scenarios involving massive sterile neutrinos. If the need for hadronic-level non-perturbative inputs is sufficiently motivated within a given scenario, the lattice-QCD community will invest in providing the values of the relevant matrix elements.

\subsection{Extracting Physical Observables}\label{sec:extracting}
The conventional paradigm in lattice QCD consists of Monte Carlo sampling of vacuum gauge-field configurations, then obtaining quantum-mechanical expectation values as a statistical average of the value of the operators over these configurations. The method, therefore, relies on Wick rotating the QCD action to Euclidean spacetime to allow a real positive weight for importance sampling. Consequently, the connection to real-time observables such as scattering and transition amplitudes becomes obscure as no notion of asymptotic states is present in imaginary time. Lattice-QCD calculations, however, are performed in a finite (and discrete) volume to make otherwise infinite-dimensional path integrals finite. It turns out that the finite-volume energy spectrum encodes information about the interactions, and via a mapping, gives access to two-hadron scattering amplitudes. The low-lying spectrum is accessible from lattice-QCD calculations of two-point correlation functions of the two-hadron state. This formalism, known as L\"uscher's method~\cite{Luscher:1986pf,Luscher:1990ux}, has been extended to more general scenarios, including to three-hadron scattering amplitudes, see Ref.~\cite{Briceno:2017max,Hansen:2019nir} for recent reviews. Furthermore, one-to-two hadronic transitions induced by a local current can be determined from a corresponding lattice-QCD three-point function involving the current and hadronic states, with successful applications in constraining matrix elements of relevance to flavor physics~\cite{Aoki:2021kgd,Wingate:2021ycr,USQCD:2019hyg}. The generalization of this formalism, known as Lellouch-L\"uscher method~\cite{Lellouch:2000pv}, is essential in determining the $nn \to pp$ transition amplitude from lattice QCD (see Refs.~\cite{Detmold:2004qn,Briceno:2012yi} for early formalisms for two-nucleon transition amplitudes).
\begin{figure}[!t]
    \centering
    \includegraphics[scale=0.6]{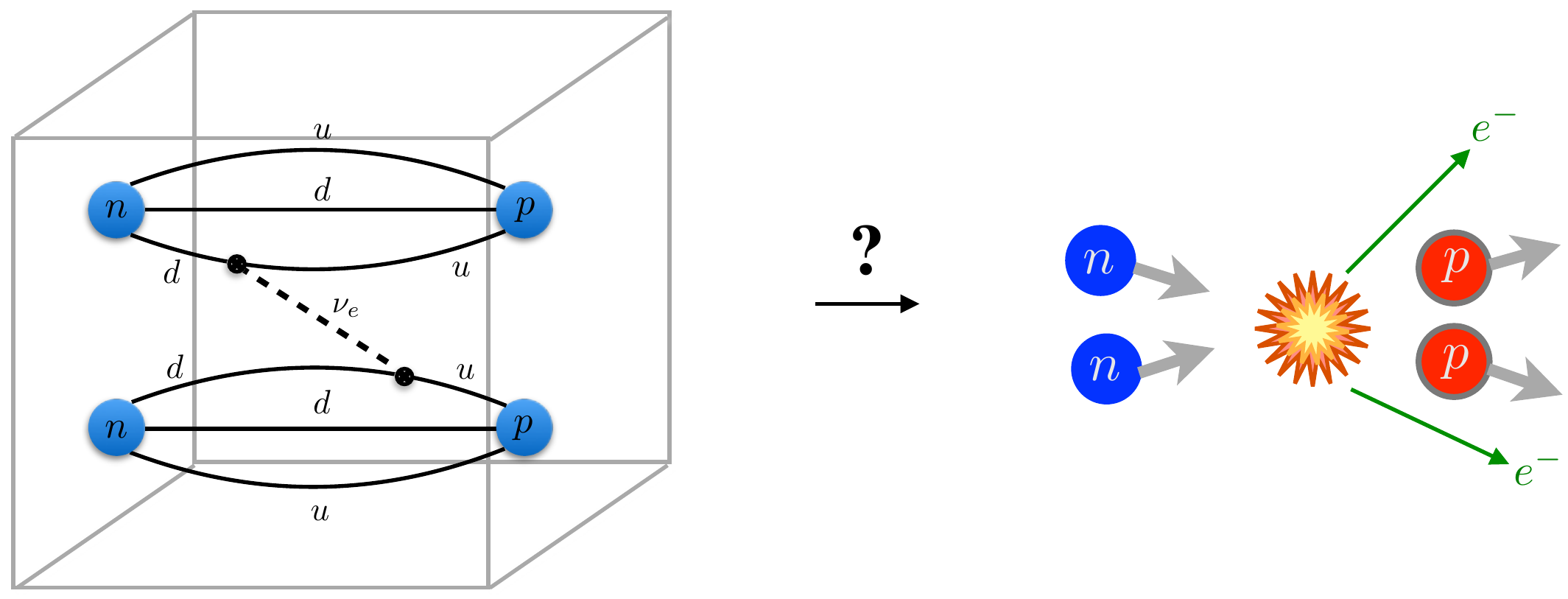}
    \caption{One of the steps involved in lattice-QCD calculations of relevance to the $0\nu\beta\beta$ program is to determine how the physical transition rates can be accessed from a lattice-QCD calculation that is performed in a finite and Euclidean spacetime. This process must be done within each LNV scenario and may need the EFT descriptions to be assisting the matching. Figure is taken from Ref.~\cite{Davoudi19}.}
    \label{fig:LQCD-matching}
\end{figure}

In particular, general model-independent formalisms for accessing one-to-two and two-to-two hadronic transition amplitudes induced by local currents exist~\cite{Briceno:2014uqa,Briceno:2015csa, Briceno:2015tza}. Therefore, once lattice QCD determines the three-point functions relevant for the $nn \to pp$ process with the higher-dimensional local operators introduced in Sec.~\ref{sec:EFTs-for-LNV}, these can be turned into the physical two-nucleon matrix elements of interest. Similarly, if the matrix elements of relevance to the subprocess $n \to \pi p$ are needed to constrain the hadronic EFTs, the path to evaluating such matrix elements is clear. The challenge to be faced in the upcoming years is to not only accurately and precisely determine the relevant lattice-QCD matrix elements, but also to constrain two-nucleon elastic scattering amplitudes at the quark masses at which the $nn \to pp$ calculations are performed. This is because the Lellouch-L\"uscher-type matching conditions require information on the energy dependence of the two-nucleon scattering amplitude near the transition energy. This puts further emphasis on reliable and precise two-nucleon spectroscopy from lattice QCD, as described in Sec.~\ref{sec:NN-LQCD}.

Furthermore, for the scenario involving a light Majorana neutrino, matrix elements of two spacetime-separated insertions of local currents are required, where hadronic and leptonic contributions are convoluted via a neutrino propagator, hence complicating the matching process. In fact, due to the long-range nature of the light-neutrino propagation between nuclear states, the separation of short- and long-distance effects necessary for arriving at a general model-independent mapping is obscured, but the matching can be perfectly developed within a corresponding EFT. Building upon the matching formalisms for one-to-one and simpler two-to-two bi-local matrix elements~\cite{Christ:2015pwa,Briceno:2019opb, Feng:2020nqj,Davoudi:2020xdv}, the formalism for matching to the leading-order pionless EFT for the $0\nu\beta\beta$ decay has been recently developed~\cite{Davoudi:2020gxs}, hence providing the path to constraining the leading-order unknown short-distance LEC $g_\nu^{NN}$ introduced in Sec.~\ref{sec:EFTs-for-LNV}. With a non-local matrix element, another involved feature is the possibility of intermediate multi-hadron states with on-shell kinematics, which give rise to a different analytic structure of the four-point function in Euclidean vs. Minkowski spacetime~\cite{Christ:2015pwa,Briceno:2019opb, Feng:2020nqj,Davoudi:2020xdv,Davoudi:2020gxs}, adding another layer of complexity to applying the matching condition in practice. Explicitly, not only the two-nucleon scattering amplitudes need to be determined consistently in the calculation but also the three-point functions characterizing transitions from initial to intermediate and intermediate to final states need to be determined in the relevant kinematic range. Nonetheless, for the simplest $nn \to pp$ process at accessible volumes in future lattice-QCD calculations, no such on-shell intermediate state is expected, as long as the neutrino propagation is properly IR-regulated in the lattice-QCD calculation~\cite{Davoudi:2020gxs}, e.g., by removing its zero mode in the zero-mass limit, see Sec.~\ref{sec:operators-and-matrix-element-techniques}.

The application of this matching condition to synthetic data~\cite{Davoudi:2021noh} suggests that future lattice-QCD computations of the $nn \to pp$ matrix element and two-nucleon energies at physical values of the quark masses need to reach below-10$\%$ (total) uncertainty to provide a first-principle QCD determination of the $g_\nu^{NN}$ LEC with similar or smaller uncertainty compared with the existing indirect estimate~\cite{Cirigliano:2020dmx,Cirigliano:2021qko}. This defines a precision goal for lattice-QCD studies of $0\nu\beta\beta$ decay in the light Majorana-neutrino scenario over the next decade.

\subsection{Computational-Resource Requirement and Future Directions}
A (near) physical-point determination of relevant matrix elements for the $nn \to pp$ process for both the light neutrino-exchange and high-scale scenarios is a challenging but likely achievable goal over the next decade, provided that sufficient computational resources are invested in reaching required accuracy and precision in two-nucleon observables from lattice QCD. While estimates for this need as a function of quark masses from a statistical standpoint may be derived using the signal-to-noise argument discussed in Sec.~\ref{sec:NN-LQCD}, the true computational cost will depend largely on a comprehensive study of the systematics of two-nucleon scattering near the physical quark masses. This study will constitute a major part of the effort in this program over the next few years, and will inform the projection of what can be achieved for the $0\nu\beta\beta$ program over the next decade.

One can envision the next stage of developments once the earlier, yet challenging goals identified above, are achieved. For example, since the output of the lattice-QCD calculation is the matrix element for the $nn \to pp$ process directly from QCD, it is independent of the model or EFT with which one may choose to describe the process. This means that the lattice-QCD matrix element can be matched to an EFT even at higher orders. However, with only one matrix element, the leading-order LECs and the new LECs occurring at next-to-leading orders cannot be constrained simultaneously. These include, for example, $n\pi p$ and $\pi\pi$ LECs. Fortunately, these can be separately constrained by dedicated lattice-QCD calculations of the underlying subprocesses as described Sec.~\ref{sec:operators-and-matrix-element-techniques}. Consequently, combining these lattice-QCD outputs for various contributions can enable a test of the proposed power counting in the EFT, and shed light on the true significance of each contribution. Matching the $nn \to pp$ matrix element to the EFT beyond the leading order, nonetheless, requires an extension of the mapping between Euclidean finite-volume correlation function and Minkowski infinite-volume amplitude as described in Sec.~\ref{sec:extracting}. This will likely be more involved but can, in principle, be derived by extending the leading-order formalism.

Furthermore, lattice QCD is capable of accounting for a massive neutrino, hence will be able to allow for constraints on the sterile-neutrino scenarios. Neutrino mass is a free parameter in the lattice-QCD calculations described in Sec.~\ref{sec:operators-and-matrix-element-techniques}, as the neutrino propagator is implemented as a numerical function, and neutrinos are not dynamical degrees of freedom in the simulation. The matching formalism for accessing physical scattering amplitude or in turn the LECs of the EFT will need to be modified to incorporate a non-zero mass for the neutrino. Nonetheless, such extensions are straightforward considering the prior developments in including IR-regulated QED interactions in lattice-QCD calculations via introducing a non-zero photon mass~\cite{Endres:2015gda}.

Finally, lattice QCD can, in principle, enable constraints on multi-nucleon contributions to the rate of $nn \to pp$ transitions in light nuclei. Although transitions in light nuclei are not pertinent to experiment, they can still provide valuable benchmarks to quantify the role of higher-body effects in the transition matrix elements. Such an effort has recently started using \emph{ab initio} nuclear many-body methods~\cite{Pastore:2017ofx,Basili:2019gvn,Yao:2020mw}, and lattice-QCD constraints on the same matrix elements will provide further tests of the EFTs and the methods used. First attempts at lattice-QCD calculations at large quark masses for low-lying spectra, $\beta$ decay rates, and several structure properties of light nuclei have been performed with up to $A = 4$ (see Ref.~\cite{Davoudi:2020ngi} for a recent review). Nonetheless, these calculations exhibit unquantified systematic uncertainties that can only be controlled by lowering the quark masses toward the physical point, controlling potential excited-state contaminations, and discretization effects. The computational resources required to properly control these effects will likely be an order of magnitude larger than those needed for two-nucleon calculations, based on the signal-to-noise arguments presented in Sec.~\ref{sec:NN-LQCD}. Besides the need for generating precise and controlled outputs from lattice QCD in the upcoming decades for higher-body matrix elements, the corresponding matching formalisms need to be put in place, which marks another challenging direction of formal research in this problem. Alternatively, lattice-QCD and \emph{ab initio} few-body calculations can be matched directly in a finite volume~\cite{Barnea:2013uqa,Kirscher:2017fqc,Detmold:2021oro,Detmold:2004qn}. Nonetheless, non-trivial analytic continuation of the Euclidean matrix elements from lattice QCD in the long-range scenario need to be accounted for.

\section{Nuclear Structure and the Computation of Double-Beta Transition Matrix Elements}
\noindent
The last row in the ``tower of EFTs" displayed in Figure \ref{fig:landscape} refers to the many-body methods that ultimately produce the NMEs needed to interpret experiments.  As was discussed, the nucleon operators used in the matrix elements depend on the the decay mechanism.  Over the past 30 years, significant work has gone into computing the NME associated with light-neutrino exchange.  Most results have come from phenomenological models but, in the last few years, \emph{ab initio} calculations have begun to appear.  We discuss both of these approaches below, with an emphasis on the latter, which provides the best hope for accurate results with quantifiable uncertainty.

\subsection{Current Status}
\begin{figure}[t]
    \centering
    \includegraphics[width=.8\textwidth]{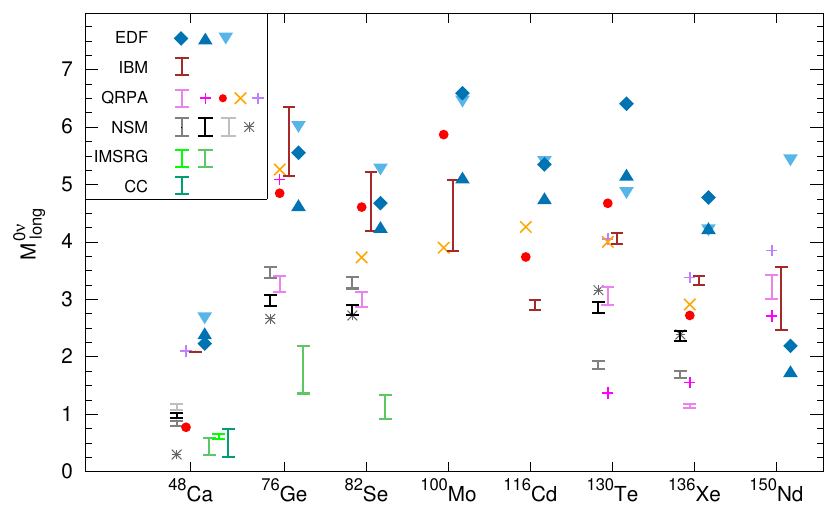}
    \caption{Results from many models for the long-range piece of the light-neutrino-exchange NMEs.  Figure is taken from Ref.~\cite{Agostini:2022zub}.}
    \label{fig:spread}
\end{figure}

``Phenomenological models'' are meant to refer to situations that contain parameters in Hamiltonians and other operators that are fit to data in nuclei with many nucleons, typically a number close to that in the nuclei one wants to understand.  The models include~\cite{Engel:2017aq} the shell model, the quasiparticle random-phase approximation (QRPA) both within energy-density functional (EDF) theory and with more schematic ingredients, the interacting boson model (IBM), and the generator coordinate method (GCM) within EDF theory. Such models work best when the parameters can be adjusted so as to fit the desired observable in nuclei near the ones one cares about.  The problem with applying them to $0\nu\beta\beta$ NMEs is that no NME data exist at all.  Without such data, it is difficult to estimate the uncertainty in a phenomenological result. The problem results in a spread of NMEs from many models---shown for the long-range part of the light neutrino-exchange mechanism in Fig.~\ref{fig:spread}---with no justifiable way to determine where within (or above or below) that range the true matrix elements lie.

These difficulties have recently motivated the development and
application of  non-perturbative \textit{ab initio} 
many-body methods to double-beta decay.  Such methods start with 
interactions and operators 
determined from QCD and/or fit to data in very light nuclei ($A=2,
3$, or 4), and then produce solutions to the Schr\"odinger equation in heavier nuclei, with systematically improvable approximations.  Three distinct \textit{ab initio} methods have been applied to the heavy open-shell nuclei of interest to experimentalists.  The first two are variants of the In-Medium Similarity Renormalization Group (IMSRG), an approach in which one uses renormalization-group flow equations to decouple a predefined ``reference'' state, ensemble, or subspace from the bulk of the many-body Hilbert space.  After this approximate diagonalization via flow equations, familiar methods can be used to complete the job.  The more the predefined space contains the low-lying states of interest, the better the procedure, which is
implemented only approximately, works.  

The first IMSRG variant used to compute $0 \nu \beta \beta$ matrix elements is the \textit{In-Medium Generator-Coordinate Method (IM-GCM)}~\cite{Yao:2020mw,Yao:2021bs,Wirth:2021pij}: In closed-shell nuclei, a ``reference'' Slater determinant will suffice as an approximation but in more complicated nuclei, one needs a state or space that builds in some low-energy correlations.  The IM-GCM uses the generator coordinate method (GCM), which superposes symmetry-violating mean-field states with a range of deformations,
pairing gaps, etc., and projects out pieces of each that conserve angular momentum and particle number, to construct a reference state that contains the important collective correlations, allowing the renormalization-group equation to incorporate the physics that it treats best---non-collective excitations---into an effective interaction and double-beta decay operator.  After the flow equations are solved, one diagonalizes the resulting Hamiltonian in a space consisting of the already mostly decoupled GCM reference state and other GCM states with different collective properties. Then, the effective $0\nu \beta\beta$ operator is used to compute the transition between this state and a similar one in the other nucleus. 
The second variant is the \textit{Valence Space IMSRG} (VS-IMSRG)~\cite{Belley:2021ek}.   Here, the flow equations are constructed to decouple a valence shell-model space.  After solving them, one diagonalizes the resulting effective Hamiltonian in the initial and final nuclei and uses the effective $0\nu \beta \beta$ operator to compute the transition matrix element.  

A third approach, \textit{Coupled Cluster (CC) Theory}, has also been applied~\cite{Novario:2021fc}.  It starts with an exponential ansatz for the ground state of an even-even nucleus: $\ket{\Phi} = e^{T} \ket{\text{HF}}$, where $\ket{HF}$ is the Hartree-Fock Slater determinant and $T$ contains operators that create particles and holes. Even if one truncates $T$ to include only few-particle--few-hole excitations, one gets a good approximation to the ground state because the exponentiated operator
creates many-particle--many-hole excitations.  To apply this approach to $0\nu\beta\beta$ decay, one needs to represent the ground state of one of the two nuclei in the decay as a two-proton-particle--two-neutron-hole (or vice versa) excitation of the other, with additional neutron-neutron and proton-proton particle-hole corrections.  A version of the formalism that allows deformed states is necessary in most $0\nu\beta\beta$ candidate nuclei~\cite{Novario:2020gr,Novario:2021fc}, but that also entails the eventual restoration of the rotational symmetry. In practical applications, an (approximate) symmetry restoration in CC has only recently been achieved for the first time~\cite{Hagen:2022tqp}.

Starting from the same two- plus three-nucleon interactions and using the same transition operator, these methods have been used to produce NMEs for the lightest ``realistic'' candidate nucleus, $\nuc{Ca}{48}$. Within roughly estimated uncertainties for each method, the results were found to be mutually compatible~\cite{Yao:2020mw,Belley:2021ek,Novario:2021fc} (cf.~Fig.~\ref{fig:spread}). The methods were also successfully benchmarked against each other as well as exact diagonalization approaches in light nuclei~\cite{Basili:2019gvn,Novario:2021fc,Yao:2021bs}. The VS-IMSRG approach was also used in a first attempt to calculate the NME for $\nuc{Ge}{76}$~\cite{Belley:2021ek}, but much work needs to be done in this approach as well as in IM-GCM and CC to properly assess all theoretical uncertainties in the coming years.

\subsection{Next Steps}

Phenomenological models can continue to improve.  Density functionals are becoming more accurate~\cite{McDonnell13,NavarroPerez18}, and are beginning to incorporate ingredients such as two-body currents from chiral EFT~\cite{Ney21}, which will lead to better QRPA calculations.  The shell model is able to include ever-larger valence spaces~\cite{Shimizu17} and short-range correlations extracted from Quantum Monte Carlo calculations in light nuclei~\cite{Weiss21}.  New degrees of freedom can be added to the IBM.  Such developments cannot help but yield more accurate NMEs.  Though it may never be possible to assign a meaningful uncertainty to the results, they should be vigorously pursued.

Nonetheless, the best opportunity for significant improvement is on the \emph{ab initio} front.  All three of the methods that have 
produced the first wave of \emph{ab initio} NMEs rely on approximations that allow  systematic improvements:
\begin{itemize}[leftmargin=*]
    \item[$\circ$] In both IMSRG approaches, the flow equations are formulated in a basis of up to two-body operators that are normal ordered with respect to the reference state. The size  of that basis scales as $\mathcal{O}(N^4)$ and the cost for evaluating the flow equations is $\mathcal{O}(N^6)$, where $N$ is the dimension of the single-particle Hilbert space~\cite{Hergert:2016jk,Hergert:2017kx}. 
    As in any renormalization-group approach, one can systematically improve the calculation by ``naively'' increasing the size of the operator basis, or by improving the basis operators that are used to represent the renormalization-group flow. In the language of the IMSRG, the former implies keeping up to three-body operators, which would increase the memory requirements to $\mathcal{O}(N^6)$ and the naive computational cost to $\mathcal{O}(N^9)$. 
    The latter can be achieved by selecting reference states that better capture the physics of the evolving system, avoiding the scaling problem. While such a selection scheme could introduce a bias, the IMSRG offers a powerful diagnostic tool: If the operator bases are sufficiently complete or optimized, the evolution should be unitary and therefore observable quantities should be independent of the continuous parameter $s$ that parameterizes the IMSRG flow\footnote{There are similarities and possibly connections to gradient flows in Quantum Field Theory.}
   ~\cite{Hergert:2016jk,Gebrerufael:2017fk,Frosini:2021ddm}.
    In the IM-GCM, one uses knowledge of collective properties to choose a (small) basis of configurations to diagonalize the Hamiltonian. As in the case of the IMSRG operator basis, the flow parameter dependence can serve as a tool for assessing whether the basis is sufficiently complete~\cite{Frosini:2021ddm}.
    In the VS-IMSRG, the effective Hamiltonian and operators are defined for a particularly chosen valence space (or active space) consisting of $A_v < A$ nucleons, while the remaining nucleons remain inert in a ``frozen core''~\cite{Stroberg:2019th}. In principle, one can probe the uncertainties due to that assumption by taking the no-core limit and making all nucleons active. This will greatly increase the cost of the numerical diagonalization, even with IMSRG pre-processing, and will likely put nuclei such as $\nuc{Ge}{76}$ out of reach~\cite{Gebrerufael:2017fk}, but one can likely find a middle ground and again rely on the flow diagnostics for uncertainty assessment.
    
    \item[$\circ$] In CC theory, the main approximations are the truncations of the cluster operator and the excitation operator that maps  either the parent or (grand)daughter nucleus' ground state to the other required ground state. Current calculations typically retain up to two-nucleon terms, or ``doubles'' excitations in either operator, at a (naive) cost of $\mathcal{O}(N^6)$, similar to that of the standard IMSRG. In Ref.~\cite{Novario:2021fc}, three-nucleon (triples) contributions to the cluster operator are approximately accounted for, but a more comprehensive treatment may be required to achieve the desired precision for the NMEs. Because the NME calculations appear to require deformed bases, work to restore the broken rotational symmetry will be important as well~\cite{Hagen:2022tqp}.
\end{itemize}

\begin{figure}[t]
  \centering
  \setlength{\unitlength}{\textwidth}
  \includegraphics[width=0.5\unitlength]{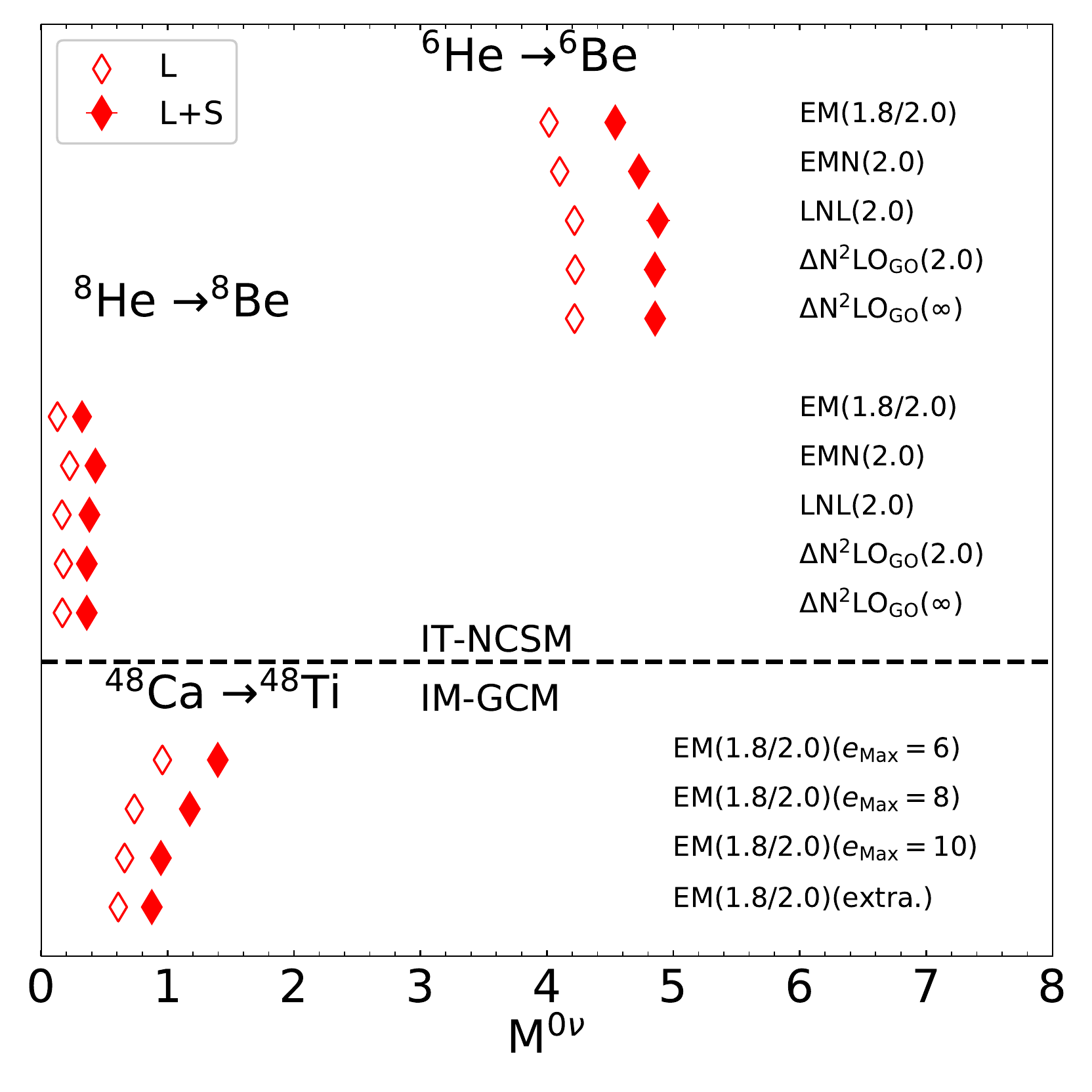}
  \vspace{-5pt}
  \caption{\label{fig:NME_contact}%
   NMEs $M^{0\nu}$ for the benchmark nuclei $\nuc{He}{6}$ and $\nuc{He}{8}$ as well as the realistic candidate $\nuc{Ca}{48}$, calculated using IM-GCM and exact diagonalization (IT-NCSM) with different chiral nuclear forces and both long-range only and long- plus short-range transition operators (see Sec.~\ref{sec:EFTs-for-LNV} and Refs.~\cite{Cirigliano:2018hja,Cirigliano:2019vdj,Cirigliano:2020dmx}). Figure is taken from Ref.~\cite{Wirth:2021pij}. 
  }
\end{figure}

All the improvements mentioned above are underway, and should be ready for producing next-generation NMEs in the near-term future.
In addition to developments in many-body theory, however, advances in the EFT-based inputs are needed. To improve on the first wave of NMEs, the nuclear interactions and $0\nu\beta\beta$ transition operators should be constructed at the same (chiral) EFT order, and the same regulators and cutoff values should be used. The LECs for the interactions and the transition operator can be determined independently at any chosen order and for any regularization scheme. In the case of the transition operator, doing so entails the inclusion of the two- and three-body operators from the chiral EFT treatment of $0\nu\beta\beta$ decay~\cite{Cirigliano:2018hja,Cirigliano:2019vdj,Cirigliano:2020dmx}).  New short-range terms will be required for renormalization, like the recently discovered leading-order contact term  (see Sec.~\ref{sec:EFTs-for-LNV} \cite{Wirth:2021pij}). Figure~\ref{fig:NME_contact} shows that this last contribution leads to a robust and significant enhancement of the NMEs in light benchmark nuclei as well as in $\nuc{Ca}{48}$, but this analysis will have to be repeated with consistently constructed interactions and operators. 

Because the LECs of nuclear interactions are determined by fitting data in two- and three-nucleon systems, one naturally has to ask to what extent the uncertainties in these data will affect the precision of predicted NMEs. Propagating the uncertainties in 20-30 parameters through a complex and costly many-body calculation is a daunting task, but significant progress has come from systematic Bayesian methods (see, e.g., Ref.~\cite{Phillips:2021ep} and references therein). The integration of these techniques into nuclear many-body workflows can offer diagnostic insights into the performance of the underlying chiral EFT, e.g., whether the breakdown scale has been properly identified or whether the power counting works as expected~\cite{Furnstahl:2015bl,Melendez:2017fj,Wesolowski:2016ek,Melendez:2019ax,Wesolowski:2019zv,Wesolowski:2021vr}. Coupling Bayesian methods with efficient emulators that reduce the computational cost for parameter sweeps and sensitivity analysis by many orders of magnitude~\cite{Frame:2018bv,Konig:2020fn,Ekstrom:2019tw,Furnstahl:2020yf,Wesolowski:2021vr} makes uncertainty propagation much more manageable, but challenges remain for the community in the coming years. Perhaps chief among them is that the emulators are only as good as their training data: They cannot recover physical effects that are explicitly excluded by truncation, and the samples of parameter space must be comprehensive enough to pin down the important features of the interactions.

It is clear from this discussion that the computation of next-generation NMEs for 
$0\nu\beta\beta$ candidate nuclei will require considerable amounts of computing 
time as well as investments into the development of many-body codes to ensure that 
the allocated time is used efficiently.  Although CC calculations have been 
successfully scaled to leadership-class supercomputers like Summit, work is required to do the 
same for IMSRG evolution and the GCM that generates reference states. In particular, the codes must be extended to 
leverage accelerators such as GPUs.  For the first-wave NME calculations, the efforts 
of several groups were combined to construct transition-operator matrix elements 
that serve as input for the aforementioned many-body methods. This ``pipeline'' 
should be optimized and strengthened in order to reduce the turnaround time between the construction of the 
operators in a particular scheme and their application in NME calculations.

\subsection{Future Directions}
The main task of nuclear-structure theory for $0\nu\beta\beta$ decay is the computation of the ground-state wave functions for the initial and final nuclei. In certain cases, excited states with specific quantum numbers and energies may be required as well, e.g., to test the quality of the closure approximation for the light-Majorana neutrino $0\nu\beta\beta$ operator~
(see, e.g., \cite{Engel:2017aq} and references therein).

The wave-function computations need to be performed only once for a particular scheme, i.e., with interactions at a given order of chiral EFT, a particular cutoff and regulator type, and a particular set of interaction LECs. This fact offers flexibility to explore $0\nu\beta\beta$ mechanisms other than light-Majorana neutrino exchange (see Sec.~\ref{sec:EFTs-for-LNV}), provided that the transition operator is compatible with the chosen regulators and does not probe high momenta. An example (though technically still for light-neutrino exchange) is the recent exploration of the leading short-range contribution to the $0\nu\beta\beta$ transition operator in Ref.~\cite{Wirth:2021pij}, for which existing wave functions were reused (with results shown in Fig.~\ref{fig:NME_contact}).
    
Ultimately, persistent questions about the implementation of EFT interactions and operators in traditional many-body methods should be addressed.  Among these are the following:
    \begin{itemize}[leftmargin=*]
        \item[$\circ$] Can the power counting of chiral EFT be made rigorous within a non-perturbative many-body calculation (see, e.g., \cite{Kolck:2020dn,Epelbaum:2020rr,Phillips:2022uc,Griesshammer:2021zzz} and references therein)?  
        As an example, consider the following issue:  In the usual Weinberg power counting, chiral EFT amplitudes are iterated to all orders to generate nuclear potentials because the $\pi N$ coupling is not perturbative. At present, this procedure is carried out in all partial waves, but it has been argued that it should be applied only in certain partial waves (see Ref.~\cite{Kolck:2020dn} and references therein).
        To employ the resulting interactions in a many-body system, one should then solve the Schr\"odinger equation with operators from the iterated partial waves to construct a non-perturbative wave function,  and treat the remaining terms perturbatively.  Such an approach may now be possible because of the recent progress in the use of correlated reference states for CC and/or IMSRG as well as the implementation of high-order perturbation theory for nuclei and nuclear matter (see Ref.~\cite{Drischler:2019xuo,Tichai:2020ft} and references therein).
        \item[$\circ$] How important are $\Delta$ resonances?  The $N-\Delta$ mass difference is small compared to the breakdown scale of chiral EFT. Virtual $\Delta$ degrees of freedom have been included to some extent in the construction of recent chiral potentials (see Refs.~\cite{Piarulli:2018xi,Piarulli:2020dp,Jiang:2020bj,Epelbaum:2020rr} and references therein) and the study of electromagnetic currents~\cite{Pastore:2011dq,Piarulli:2013vn,Schiavilla:2019to,Krebs:2020te}  but the systematic exploration of $\Delta$-full chiral EFT in many-body applications is far from complete. In a $\Delta$-full EFT, certain contributions are promoted in the power counting: For example, the leading three-nucleon force appears at NLO instead of NNLO. This is bound to have an impact on the convergence behavior of observables.
        \item[$\circ$] What are the effects of varying regulators?  Quantum Monte Carlo calculations with a local chiral three-nucleon interaction found that operator bases that are connected by Fierz transformations and therefore equivalent did \emph{not} yield compatible results for light nuclei due to regulator artifacts~\cite{Lynn:2016ec,Lynn:2017eu}. This issue and similar problems have prompted closer investigations of local, semi-local, or completely nonlocal regularization schemes in chiral EFT, and how to implement them consistently (see Ref.~\cite{Epelbaum:2020rr} and references therein).  
        \item[$\circ$] How useful are statistical methods in resolving these problems?  As mentioned above, the  Bayesian framework developed in Refs.~\cite{Furnstahl:2015bl,Melendez:2017fj,Wesolowski:2016ek,Melendez:2019ax,Wesolowski:2019zv,Wesolowski:2021vr} can offer valuable diagnostics for the performance of EFTs.
        \item[$\circ$] 
        Finally, and only partly tongue-in-cheek: How ``effective" is an EFT that forces us to go to N${}^3$LO or N${}^4$LO to achieve an accuracy of a few percent?
    \end{itemize}

\section{Conclusions}
\noindent
With the prospect of tonne-scale $0\nu\beta\beta$ experiments, LHC searches, and improved cosmological constraints over the next decade, understanding the implications of new limits or potential discoveries for new-physics scenarios, as well as delivering $0\nu\beta\beta$ rates with minimal model dependence and quantifiable theoretical uncertainties, will continue to demand an extensive theoretical effort in the U.S. and worldwide. This white paper contains a discussion of the next theoretical directions the community should take in the context of the recent progress in the field. The conclusions of the white paper can be summarized as follows:
\begin{itemize}[leftmargin=*]
    \item[$\circ$] The observation of $0\nu\beta\beta$ decay is a smoking-gun signature of Majorana neutrinos and thus provides an essential part of our quest to understand the neutrino-mass mechanism. 
    Additional particles introduced in popular mechanisms such as the type I seesaw can affect the interpretation of $0\nu\beta\beta$ results.  
    \item[$\circ$] Some lepton-flavor models, introduced to explain neutrino-mixing parameters, also make predictions for the Majorana phases or even the absolute neutrino-mass scale. The predictions in turn lead to predictions for $|m_{\beta\beta}|$ that let $0\nu\beta\beta$ experiments help us solve the lepton-flavor puzzle.
     \item[$\circ$] A discovery of LNV in $0\nu\beta\beta$ experiments would imply that neutrinos are Majorana particles, but the exchange of such particles might not be the leading contribution to the decay. 
     Collider searches, particularly at the LHC, will play an important role in excluding or discovering other contributions from TeV-scale LNV. 
    \item[$\circ$] 
    In the early Universe, LNV at the TeV-scale can ``wash out'' any pre-existing lepton asymmetry, undermining the standard leptogenesis-to-baryogenesis mechanism. By discovering TeV-scale LNV, LHC experiments can therefore help to falsify high-scale models of leptogenesis. 
    \item[$\circ$] Simplified models of LNV at the TeV-scale provide useful schemes for investigating correlated signals of LNV across multiple kinds of experiments, spanning wide energy and distance scales.
    \item[$\circ$] The question of whether $0\nu\beta\beta$ experiments are more sensitive to TeV-scale LNV than LHC experiments does not have a model-independent answer~\cite{Cai:2017mow}. LHC experiments have the potential to provide tighter constraints than $0\nu\beta\beta$ experiments, as well as to explore complementary regions of parameter space. 
    \item[$\circ$] The LHC collaborations are encouraged to investigate simplified models for TeV-scale LNV and to make projections for the HL-LHC. Dedicated analyses of simplified models that lead to optimized search strategies may increase the ability of LHC experiments to probe TeV-scale LNV. 
    Optimized cut-flows, along the lines of those in
    Ref.~\cite{Peng:2015haa} as well as those reviewed in Ref.~\cite{Cai:2017mow}, significantly increase signal-to-background over standard same-sign dilepton cuts. 
    \item[$\circ$] Searches for LLPs at the LHC far detectors are encouraged to develop a program to search for LNV in tandem with searches for exotica at the main detectors~\cite{Li:2021fvw,Alimena:2019zri,Feng:2022inv,Abdullahi:2022jlv}.
    \item[$\circ$] EFTs play an indispensable role in connecting any BSM model of LNV, such as Majorana neutrinos, sterile neutrinos, or simplified models of TeV-scale LNV, to low-energy observables.
    Reducing theoretical uncertainties by 
    calculating the necessary LECs in lattice QCD,
    constructing higher-order corrections to the $0\nu\beta\beta$ transition operator in chiral EFT, and calculating NMEs with \textit{ab initio} methods is critical to future progress.
    \item[$\circ$] Over the next few years, lattice-QCD calculations will prioritize systematic studies of two-nucleon energy spectra and elastic scattering amplitudes at a range of quark masses leading toward the physical point. This is because these calculations will inform the upcoming $nn \to pp$ calculations on the most optimal choice of QCD-based interpolating operators and the quantification of systematic uncertainties. Besides, two-nucleon properties enter the determination of physical transition amplitudes for $0\nu\beta\beta$ processes. 
    \item[$\circ$] Lattice QCD promises to constrain the unknown LECs in the EFT descriptions of $0\nu\beta\beta$ decay in the two-nucleon sector, and to quantify nuclear effects when transitions occur in systems with more than two nucleons. Methods are in place, and benchmarked in simpler systems, to calculate $0\nu\beta\beta$ matrix elements from QCD. Since more than a single hadron is involved in the decay process, general mappings are developed to connect the lattice-QCD output to physical observables. Nonetheless, these mappings sometimes require an EFT to complete the matching, directly or indirectly. They need to be further extended for higher-order EFT processes or when more nucleons are involved.
    \item[$\circ$] At the nuclear-structure level, phenomenological models are being superseded by \emph{ab initio} methods, in particular  CC and the IMSRG, that make use of EFT operators.  These methods have already produced a first wave of NMEs.
    \item[$\circ$] The CC and the IMSRG methods can both still be made more accurate.  The improvement, however, will require extensive computing resources and a rewriting of codes, particularly for the IMSRG, 
    to exploit those resources.
     \item[$\circ$] The use of EFTs in conjunction with non-peturbative many-body methods such as CC and the IMSRG needs to be examined carefully.  Inconsistencies that have mostly been swept under the rug so far must be addressed if genuinely quantifiable uncertainty in NMEs is desired. 
\end{itemize}

 \acknowledgements
\noindent
We thank M.~Agostini, G.~Benato, J.~Detwiler, and J.~Menendez for providing Figs.~\ref{fig:masses} and \ref{fig:spread};
G.~Li for providing Figs.~\ref{fig:Mathusla1} and \ref{fig:Mathusla2}; J.~C.~Vasquez for providing Fig.~\ref{fig:mLRSM}; and S.~Urrutia-Quiroga for providing Fig.~\ref{fig:lepto}.
The work of VC was supported by the U.S. Department of Energy under Grant No. DE-FG02-00ER41132.
ZD acknowledges support from Alfred P. Sloan foundation, and Maryland Center for Fundamental Physics at the University of Maryland, College Park.
JE acknowledges support from the U.S. Department of Energy Office of Nuclear Physics under Contract DE-FG02-97ER41019.
XF acknowledges support from NSFC of China under Grants No. 12125501, No. 12070131001, and No. 12141501, and National Key Research and Development Program of China under No. 2020YFA0406400.
JG acknowledges support from the U.S. Department of Energy under Grant Contract DE-SC0012704.
The work of MLG is supported by the LDRD program at Los Alamos National Laboratory and by the U.S. Department of Energy, Office of High Energy Physics under Contract No. DE-AC52-06NA25396.
LG acknowledges support from the National Science Foundation, Grant PHY-1630782, and to the Heising-Simons Foundation, Grant 2017-228.
HH acknowledges support from the U.S. Department of Energy, Office of Science, Office of Nuclear Physics under awards No. DE-SC0017887 and DE-SC0018083 (NUCLEI SciDAC-4 Collaboration).
LCJ acknowledges support from the U.S. Department of Energy under grant DE-SC0010339 and DE-SC0021147.
EM is supported  by the U.S. Department of Energy through the Office of Nuclear Physics  and  the  LDRD program at Los Alamos National Laboratory. Los Alamos National Laboratory is operated by Triad National Security, LLC, for the National Nuclear Security Administration of the U.S. Department of Energy (Contract No. 89233218CNA000001).
The work of AN was supported by the National Science Foundation CAREER Program.
MJRM was supported in part under U.S. Department of Energy contract No. DE-SC0011095 and National Natural Science Foundation of China Grant No. 19Z103010239.
RR acknowledges the support of the Polska Akademia Nauk (grant agreement PAN.BFD.S.BDN. 613. 022. 2021 - PASIFIC 1, POPSICLE). This work has received funding from the European Union's Horizon 2020 research and innovation program under the Sk{\l}odowska-Curie grant agreement No.  847639 and from the Polish Ministry of Education and Science.
UvK's research was supported in part by the U.S. Department of Energy, Office of Science, Office of Nuclear Physics under award number DE-FG02-04ER41338.
The work of AWL was supported in part by the U.S. Department of Energy, Office of Science, Office of Nuclear Physics under award No. DE-AC02-05CH11231.


\bibliography{biblio}

\end{document}